\documentclass[aps,prb,twocolumn,secnumarabic,nobalancelastpage,amsmath,amssymb,
nofootinbib]{revtex4-2}

\usepackage{dcolumn}
\usepackage{bm}
\usepackage[latin1]{inputenc}
\usepackage[active]{srcltx}
\usepackage{ae,aecompl}

\usepackage{graphicx}
\usepackage{amsmath}
\usepackage{color}
\usepackage{braket}
\usepackage{xcolor}

\begin{document}

\title{Thermoelectric properties of topological chains coupled to a quantum dot}

\author{A. C. P. Lima$^{1}$}
\author{R. C. Bento Ribeiro$^{1}$}
\author{J. H. Correa$^{2}$}
\author{Fernanda Deus$^{3}$}
\author{M. S. Figueira$^{4}$}
\author{Mucio A. Continentino$^{1}$}

\email[corresponding author:]{muciocontinentino@gmail.com }
\affiliation{$^{1}$ Centro Brasileiro de Pesquisas F\'{\i}sicas, Rua Dr. Xavier Sigaud, 150, Urca 22290-180, Rio de Janeiro, RJ, Brazil}
\affiliation{$^{2}$Universidad Tecnol\'ogica Del Per\'{u}, Nathalio Sanchez, 125, Lima, Per\'{u} }
\affiliation{$^{3}$Universidade do Estado do Rio de Janeiro, Faculdade de Tecnologia, Departamento de Matem\'{a}tica, F\'{i}sica e Computa\c{c}\~{a}o, Rodovia Presidente Dutra km 298, 27537-000, Resende, RJ, Brazil}
\affiliation{$^{4}$Instituto de F\'{i}sica, Universidade Federal Fluminense, Av. Litor\^anea s/N, CEP: 24210-340, Niter\'oi, RJ, Brasil}

\begin{abstract}
Topological one-dimensional  superconductors can sustain in their extremities zero energy modes  that are protected by different kinds of symmetries. The observation of these  excitations in the form of Majorana fermions is one of the most intensive quests in condensed matter physics. In this work we are interested in another class of one dimensional topological  systems, namely topological insulators. These  also present symmetry protected end modes with robust properties and do not require the low temperatures necessary for topological superconductivity. We consider a device in the form of a single electron transistor coupled to the simplest kind of topological insulators, namely chains of atoms with hybridized $sp$ orbitals. We study the thermoelectric properties of the device   in the trivial, non-trivial topological phases and at the quantum topological transition of the chains.  We show that the electrical conductance and the Wiedemann-Franz ratio of the device at the topological transition have universal values at very low temperatures.  The conductance and thermopower of the device with diatomic $sp$-chains, at their topological transition,  give direct evidence of fractional charges in the system. The former has an anomalous low temperature behavior, attaining a universal value that is a consequence of the double degeneracy of the system due to the presence of zero energy modes.   On the other hand, the system can be tuned to exhibit high values of the thermoelectric figure of merit and the power factor at high temperatures.
\end{abstract}

\pacs{71.20.N,72.80.Vp,73.22.Pr}

\maketitle

\section{Introduction}
\label{sec1}

The origin of {\it thermoelectricity} can be traced back to the discovery of the Seebeck effect in the $19^{th}$ century. It consists in the production of electrical energy directly from heat, and its inverse, the Peltier effect, that transforms electrical energy into thermal energy. After the development of the first thermoelectric generators (TEGs) with applications in industry~\cite{Polozine2014}, these lost the competition with the dynamoelectric machines due to the high costs of their electrical energy generation.  Their technological development was interrupted for several decades. Only in the middle of the last century, due to the needs of the aerospacial and military industries, did the interest in developing new TEGs reappear. The thermoelectricity acquires some practical applications in those strategic areas after the discovery that the doped semiconductor $Bi_{2}Te_{3}$ and its alloys $Sb_{2}Te_{3}$, and $Bi_{2}Se_{3}$ \cite{Iofee57, Wright1958, Ioffe59}, present  high electric conductivities $\sigma$ and  low thermal conductivities $\kappa$. In consequence, those thermoelectric materials (TEM) exhibit at ambient temperatures a higher dimensionless thermoelectric figure of merit ($zT$) \cite{Iofee57, Polozine2014} and a high power factor (PF) and, until now, dominate the commercial industry of TEGs \cite{Witting2019}.

Recently, it was shown that the usual thermoelectric materials (TEM), like $Bi_{2}Te_{3}$, $Bi_{2}Se_{3}$, $Sb_{2}Te_{3}$, and $FeSb_{2}$ \cite{Xu2017,Xu2020,gooth2018} are also three-dimensional topological insulators exhibiting surface states with a single Dirac cone
and some of their striking properties are due to their strong spin-orbit coupling~\cite{Manchon2015,Witting2019,Ahmad2020} and their conducting surface states~\cite{Takahashi2012,Cassinelli2017}. A promising route to explore the effects of the topologically non-trivial surface states (TNSS) on the TE properties was followed in Ref.~\cite{Liang2016}, which studied thin films of $Bi_{2}Te_{3}$. The authors used first-principles calculations and Boltzmann theory to obtain $zT$ for different film thicknesses. They defined a unit (QL) of quintuple layers of the real material $Te-Bi-Te-Bi-Te$ and observed a $p-$type and $n-$type $zT \simeq 2$ peak in $QL=3$ when the system enters the topologically non-trivial regime from the trivial one. The results show a relevant enhancement of $zT$ due to the contribution of TNSS compared to the pristine form of bulk
$Bi_{2}Te_{3}$, $zT=0.4$. Another step in the direction of the use of TNSS states in real systems was obtained after the recent advances in the synthesis of $Bi_{2}Te_{3}$ thin films, which allows separating the bulk from the TNSS states in order to design quantum devices with improved thermoelectric properties \cite{Ngabonziza2022}.

The study of topological systems  is now one of the most active areas of research in condensed matter physics~\cite{Alicea_2012,Rachel2018,RevModPhys.90.015001,asboth2016short}. The  theoretical efforts to understand the properties of these systems has lead to the predictions of emergent excitations with unexpected properties that make them potentially useful for different types of applications. Among these works, Kitaev model for a p-wave superconducting chain~\cite{Kitaev2001} has played a fundamental role and many suggestions have appeared of how to realize
this model in actual physical systems. In the topological phase, the finite one-dimensional Kitaev superconducting chain presents Majorana, zero energy modes, at its  ends. 
The physical  implementation of the $p$-wave Kitaev model and the detection~\cite{PhysRevB.89.165314,buccheri2021violation,Sato_2017,RevModPhys.83.1057,Wang:2017wz,PhysRevLett.105.177002,PhysRevLett.105.077001,Mourik2012,PhysRevB.103.014513}  of the zero energy Majorana modes is a modern Graal in materials research. In this pursuit an initial major difficulty is to obtain a $p$-wave superconductor, since this is  far from being common in nature~\cite{Mizushima_2015}. Several proposals have been put forward to generate this type of pairing in a chain, mostly using proximity effects and magnetic fields~\cite{Alicea_2012}. 
%
%

Besides one-dimensional $p$-wave superconductors  there is a class of topological insulating~\cite{Shen2017,Hasan2010} chains that is much simpler and also presents protected zero energy modes at their ends.  Representative members of this class are  $sp$-chains  consisting  of atoms with hybridized $s$ and $p$ orbitals~\cite{Continentino2014c}. The mixing between $s$ and $p$ orbitals in neighboring ions is antisymmetric and this gives rise to non-trivial topological properties~\cite{Sun2012tb}, in close analogy with the antisymmetric $p$-wave paring of the Kitaev chain. Notice that the asymmetry of the mixing holds for any pair of orbitals that have angular momentum quantum numbers differing by an odd number. In spite of their symmetry  protection, the   edge modes in topological $sp$-chains have  distinct features from the Majoranas in the Kitaev chain. The former  are quasi-particles with a hybrid $sp$-character that are formed of two different types of Majoranas~\cite{Continentino2014c}.

The $sp$-chains may be easier to realize in practice then $p$-wave  superconductors. Also, they do not require the low temperatures necessary for superconductivity, to manifest their topological properties. 
A possible realization of the $sp$-chain is carbyne, the one-dimensional  allotropic form of carbon~\cite{carbyne2010,carbyne2014,carbyne2015,carbyne2016,carbyne2016b}. In this system the 2$s$ orbital hybridizes with a {\it single}  2$p$ orbital favoring a linear atomic alignment~\cite{LUSTOSA2022131886}.
A significant effort has been made in the synthesis of these materials that in principal can exist in a metallic state (cumulene) and in an insulating, broken symmetry state, with alternating single and triple bonds~\cite{carbyne2014}. 

As we show in the appendix monoatomic and diatomic  $sp$-chains can be mapped in two very well known topological chains,  the Su-Schrieffer-Heeger (SSH)~\cite{SSH,pham2021topological} and  the Rice-Mele (RM)~\cite{RiceMele,RevModPhys.91.015005}  chains, respectively. These chains have been intensively studied and their topological properties are well known. For this reason we study here the latter two models since they yield results for the thermoelectric properties similar to those of the $sp$-chains. 

This paper studies the thermoelectric properties of two semi-infinite Rice-Mele chains connected to a quantum dot. We investigate the device's electrical and thermal transport properties as a function of temperature, in the topologically non-trivial and trivial phases and at the topological transition. According to Refs.~\cite{Liang2016, Ngabonziza2022} we expect an increase of $zT$ due to topological states at the edges of these chains.

This work has the following structure: In section \ref{sec2}, we introduce the Rice-Mele model and present its topological properties. In section \ref{sec3}, we employ a method developed in Ref.~\cite{Thorpe76} to obtain the local  Green's function at the edge of the chain. This yields the {\it surface} density of states for the  Rice-Mele chain.  In section \ref{sec4}, we present the device consisting of two identical semi-infinite topological chains connected to a singly occupied quantum dot~\cite{maurer2021}, without correlations effects. We use linear response theory to define the thermoelectric coefficients.   In sections \ref{sec5}  and \ref{sec6} we calculate,  electrical and thermal conductances,  thermopower,  Wiedemann-Franz ratio,  power factor, and the dimensionless thermoelectric figure of merit of our device when the quantum dot is connected to monoatomic $sp$ or SSH chains, and  to  diatomic $sp$ or Rice-Mele chains, respectively. Notice that the figure of merit measures the usefulness of the device to produce electrical power. In section\ref{sec7n}, we present the high temperature results and finally, we conclude with a discussion of our results and the perspectives of our approach.

\section{The Rice-Mele model}
\label{sec2}

The Rice-Mele model has been used to describe polymeric chains with alternating bonds~\cite{RiceMele}. It is generally associated with fractional charges that arise due to their topological properties and it is used here to model diatomic $sp$-chains. Its Hamiltonian is given by
\begin{align}
\label{RMH}
&\mathcal{H}_{RM}= - V_1 \sum_n  c_{A,n}^\dagger c_{B,n}   - V_2 \sum_n c_{A,n+1}^\dagger c_{B,n} +\\ \nonumber
& (\epsilon_A-\mu) \sum_n  c_{A,n}^\dagger c_{A,n}  + (\epsilon_B - \mu) \sum_n  c_{B,n}^\dagger c_{B,n} + H.c., 
\end{align}
where $ c_{(A,B),n}^\dagger$ and $c_{(A,B),n}$ create and annihilate electrons on site $n$ of sub-lattice (A,B), respectively.  The hopping $V_1$ connect electrons in the same unit cell $n$, and $V_2$ those in different unit cells.  The site energies $\epsilon_{(A,B)}$ are different in sub-lattices A and B and $\mu$ is the chemical potential. For a semi-infinite chain the sum extends from $n=0$ to $n=\infty$.  
The SSH model is obtained from the RM model, Eq.~\ref{RMH},  when the site energies are taken equal zero, i.e., $\epsilon_A=\epsilon_B=0$.

The energy of the bands of the infinite, translation invariant  RM chain can be obtained transforming to momentum space and diagonalizing the Hamiltonian~\cite{Shen2017}. They are given by
\begin{eqnarray}
\tilde{\omega}_1(k)=-\tilde{\mu} +\sqrt{2 \tilde{V} \cos (k)+\tilde{V}^2+\tilde{\epsilon}^2+1} \\
\tilde{\omega}_2(k)=-\tilde{\mu} -\sqrt{2 \tilde{V} \cos (k)+\tilde{V}^2+\tilde{\epsilon}^2+1}. 
\end{eqnarray}
The extrema of the bands occur for $k=\pi$. Notice that there is always a gap between the bands, which is given by
\begin{equation}
\tilde{\Delta}=|\tilde{\omega}_1(\pi)-\tilde{\omega}_2(\pi)|=2 \sqrt{(1-\tilde{V})^2+\tilde{\epsilon}^2}.
\end{equation}
The {\it tilde} quantities are dimensionless,  normalized by the hopping $V_2$ and $\tilde{V}=V_1/V_2$. We took $\epsilon_A=-\epsilon_B=\epsilon$. In the case of the SSH model, with $\epsilon=0$, the band gap closes for $\tilde{V}=1$, at the topological transition.

The topological properties of the RM and SSH chains are well known~\cite{Shen2017,asboth2016short}. For the latter there is a non-trivial topological phase for $\tilde{V}<1$ characterized by a non-trivial winding number. For $\tilde{V}=1$ there is a topological  transition for a topologically trivial phase with $\tilde{V}>1$. In the non-trivial topological phase there are edge modes at the ends of a finite chain. These edge states decay into the bulk with a characteristic length that depends on the distance to the topological transition, $\xi=(1-\tilde{V})^{-\nu}$. At the topological transition $\xi$ diverges and the {\it surface state} spreads into the bulk~\cite{Continentino2014c}. For the SSH model the critical exponent $\nu=1$.

The topological properties of the RM model are more complex, but also well known~\cite{Shen2017,asboth2016short}. 
The topological phases can be characterized by Chern numbers~\cite{asboth2016short,Shen2017,RevModPhys.91.015005,Chern_RM,CHEN202150}, $n_C=-{\rm sgn}[\epsilon(V_2-V_1)]$, such that for $V_1=V_2$ or $\epsilon=0$ there are topological quantum phase transitions~\cite{Shen2017,Chern_RM,CHEN202150}. The phase with  $\tilde{V} < 1$ is topologically non-trivial.

\section{The surface density of states}
 \label{sec3}

In order to obtain the thermoelectric properties of our device, we need to calculate the surface density of states of the semi-infinite RM and SSH chains.  Here, we  use a  method  developed in Ref.~\cite{Thorpe76} that yields  the local Green's  functions at the edge of this chain.  This Green's function is  obtained from the self-consistent equation,
\begin{equation}
G_{00}(\omega) =  
\begin{pmatrix}
\omega+\mu-\epsilon & V_{1} & 0 \\
V_{1} & \omega+\mu+\epsilon & V_{2} \\
0 & V_{2} & G_{00}^{-1}
\end{pmatrix}_{(00)}^{-1},
\label{G000}
\end{equation}
from which, we can get the  surface density of states,
\begin{equation}
\label{rhos}
\rho(\omega)=\frac{-1}{\pi} {\rm Im} G_{00}(\omega).
\end{equation}

\subsection{SSH chains}
\label{SSHchains}

Let us start with the simpler case  of the semi-infinite SSH chain, which corresponds to the RM model with $\epsilon=0$.  From Eq.~\ref{G000}, with $\epsilon=0$, we obtain a self-consistent  problem involving a second degree algebraic equation for the local Green's function,
\begin{equation}
[V_{2}G_{00}]^{2}-2\alpha [V_{2}G_{00}] + 1=0,
\label{Seg}
\end{equation}
with
\begin{equation}
\alpha(\omega)=\frac{\omega^{2}+V_{2}^{2}-V_{1}^{2}}{2V_{2}\omega}.
\label{Alf}
\end{equation}
We consider the case of half-filled band and take $\mu=0$.
The surface Green's function can be directly obtained from Eq.~\ref{Seg}. It is given by,
\begin{equation}
\label{G00SSH}
G_{00}(\omega)=\frac{1}{2} \frac{1}{ \omega}\left[\tilde{\omega}^2-\tilde{V}^2+1\pm \sqrt{(\tilde{\omega}^2-\tilde{V}^2+1)^2-4 \tilde{\omega}^2}\right]
\end{equation}
where
$\tilde{\omega}=\omega/V_{2}$, 
and $\tilde{V}=V_1/V_2$.

\begin{figure}[htp]
  \begin{centering}
\begin{tabular}{cc}
\includegraphics[clip,width=0.38\textwidth,angle=0.0]{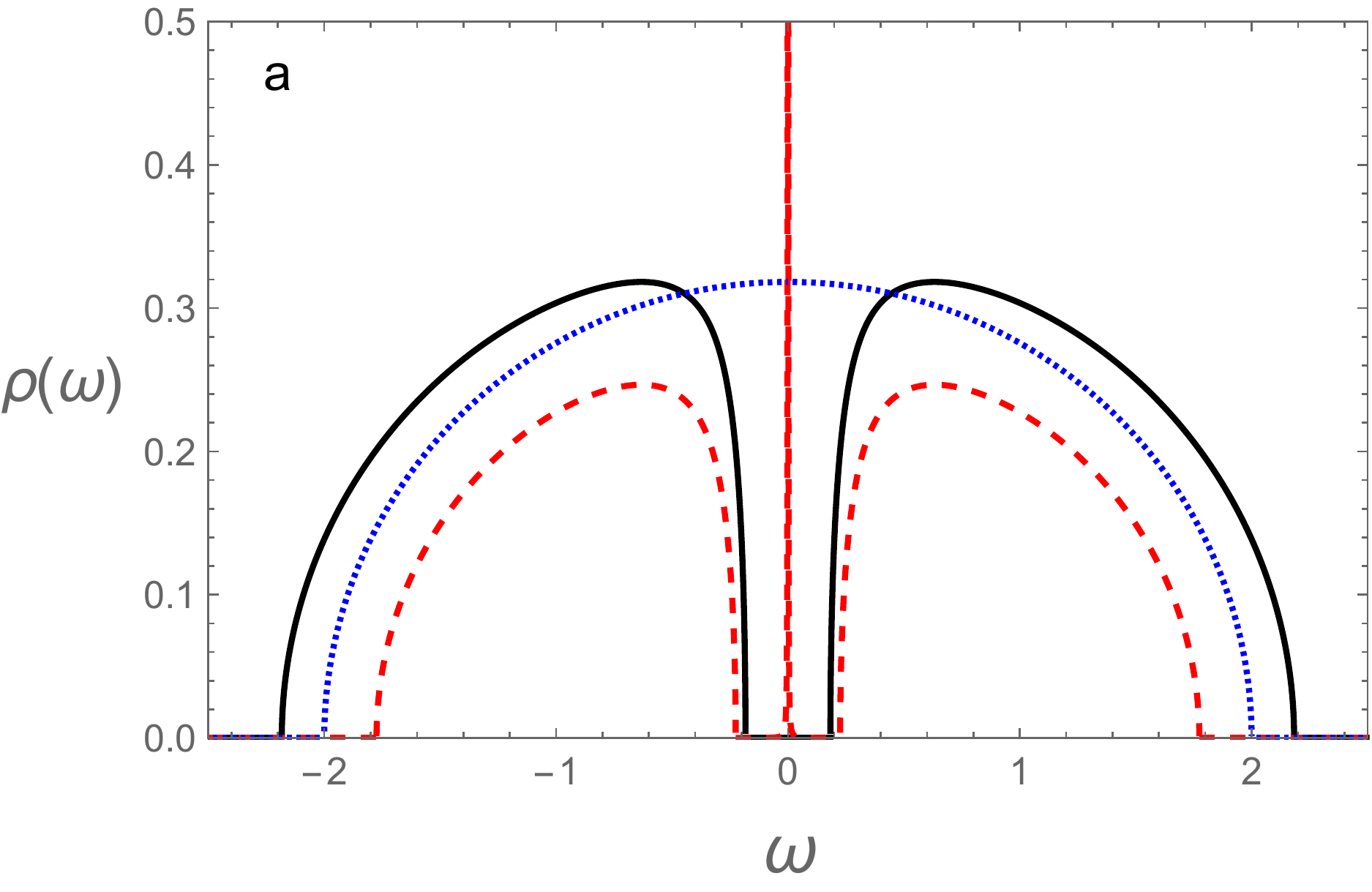}\\
 \includegraphics[clip,width=0.38\textwidth,angle=0.0]{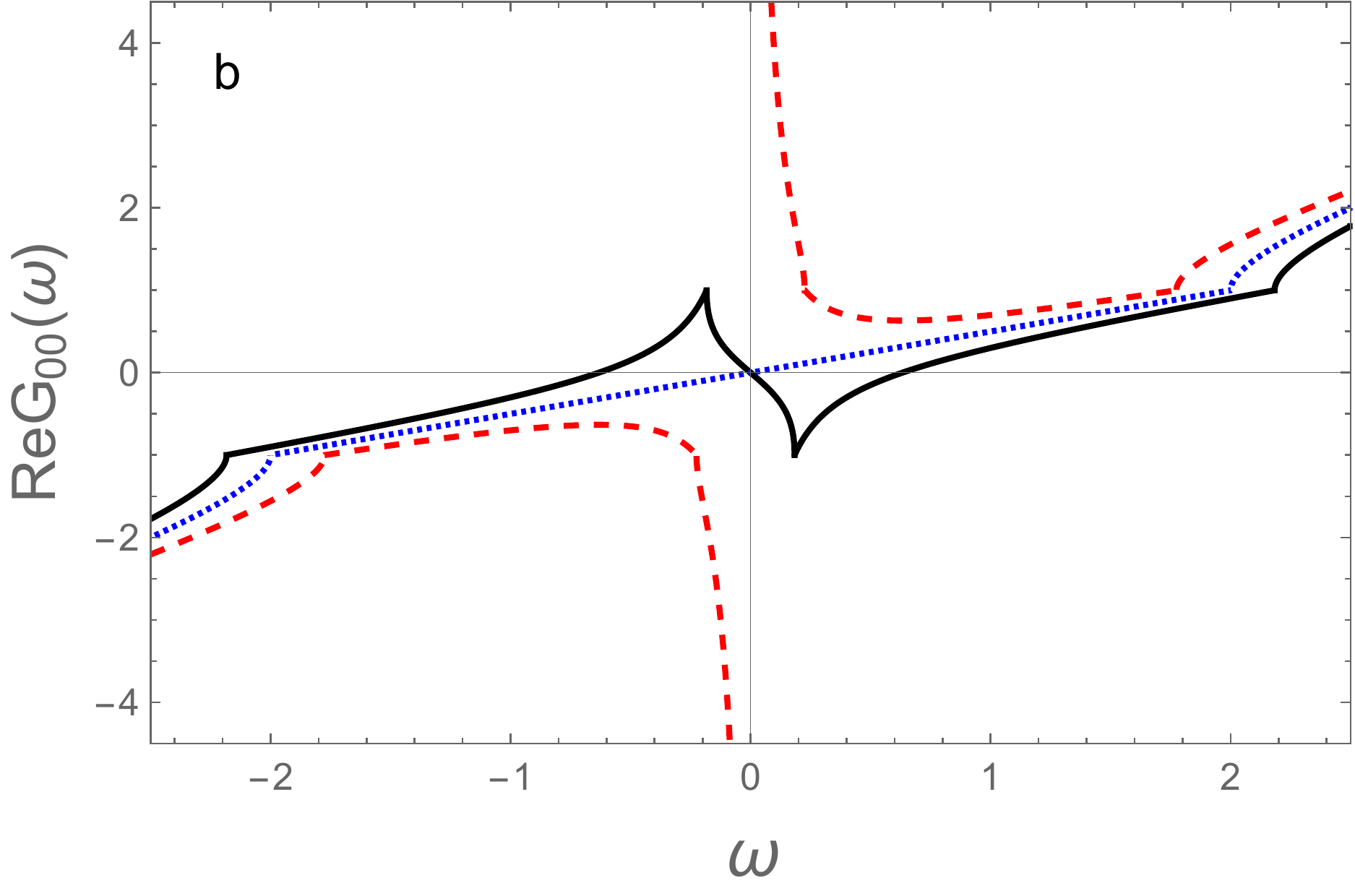} 
\end{tabular}
  \end{centering}
    \caption{  \label{fig1}(Color online)  a) Density of states at the surface of the semi-infinite SSH chain and b) real part of the surface Green's function.   In the topological phase (red, dashed),  the trivial phase (black, continuous) and at the topological transition (blue, dotted). We took $V_2=1$, such that $\tilde{\omega}=\omega$. 
 }
 \end{figure}

The surface density of states is obtained from Eqs.~\ref{rhos} and \ref{G00SSH} and is given by
\begin{equation}
\label{dens1}
\rho(\omega)= \frac{1}{2} \left\{ D \delta(\tilde{\omega}) + \frac{1}{\pi}{\rm Im} \lbrack \frac{ \sqrt{(\tilde{\omega}^2+D)^2-4 \tilde{\omega}^2}}{\tilde{\omega}}      \rbrack \right\}.
\end{equation}
where $D=1-\tilde{V}^2$ and $\tilde{\omega} \rightarrow \tilde{\omega}+i \epsilon$. The sign of the root is chosen so that the density of states is positive and from now on we take $V_2=1$. There is an additional contribution to the zero  energy mode due to the second, square root term. Considering this explicitly, we can rewrite
Eq.~\ref{dens1} as
\begin{equation}
\rho(\omega)= \frac{1}{2} \left\{ (D+|D|) \delta(\tilde{\omega}) + 
\frac{1}{\pi}\frac{{\rm Im}\sqrt{(\tilde{\omega}^2+D)^2-4 \tilde{\omega}^2}}{\tilde{\omega}}\right\},
\end{equation}
where one sees that the zero energy mode only appears for $D>0$, or $\tilde{V}<1$, i.e., in the topological phase of the chain  (see Fig. \ref{fig1}). In the trivial phase there is a cancellation and the zero energy surface mode disappears. Notice that  the zero energy mode is  a true surface state since  its energy does not coincide with any of the bulk states.
For completeness and since it will be used further on, we also  obtain the real part of the surface Green's function (see Fig. \ref{fig1}). This is given by
\begin{equation}
{\rm Re } G_{00}(\tilde{\omega})=\frac{1}{2}\left\{ D\frac{1}{\tilde{\omega}} + \tilde{\omega} +\frac{{\rm Re}\sqrt{(\tilde{\omega}^2+D)^2-4 \tilde{\omega}^2}}{\tilde{\omega}} \right\}.
\end{equation}
Notice that
$$
\lim_{\tilde{\omega} \rightarrow 0}{\rm Re } G_{00}(\tilde{\omega})=\frac{1}{2} \frac{(D+|D|)}{\tilde{\omega}}.
$$
Then,   we find that the  surface  Green's function  gives direct information on the topological state of the chain. Furthermore, the weight of the zero energy mode vanishes linearly with the distance to the topological transition ($D \propto (1-\tilde{V})$).

\subsection{Rice-Mele chains}

The surface density of states of the semi-infinite RM chain, obtained  from Eqs.~\ref{G000} and~\ref{rhos}   is given by
\begin{equation}
\label{rhorm}
\rho(\omega)\!=\!\left(D+\left| D \right| 
   \right) \delta (\omega+ \mu -\epsilon )+ 
\frac{\text{sgn}(\omega+ \mu )\Im m [R(\omega)]}{2 (\omega+ \mu -\epsilon )}
\end{equation}
with
\begin{eqnarray}
R(\omega)=\left(-(\mu +\omega )^2+(1-\tilde{V})^2+\epsilon^2\right)^{\frac{1}{2}} \times \\ \nonumber
\left(-(\mu +\omega )^2+(1+\tilde{V})^2+\epsilon^2\right)^{\frac{1}{2}}
\label{Radic}
\end{eqnarray}
Notice the presence of a surface mode at a finite energy  $\omega_S=\epsilon-\mu$, for  $\tilde{V}<1$ (since  $V_2=1$ we keep the {\it tilde } only in $\tilde {V}$).
Differently from the SSH model,   as seen in Sec.~\ref{SSHchains}, the RM  system is always gapped even at $\tilde{V}=1$.
However, the phases with $\tilde{V} > 1$ and $\tilde{V} < 1$ can still be distinguished by the absence or presence, respectively of the surface mode (besides  their Chern numbers) ~\cite{asboth2016short,Shen2017,Chern_RM,RevModPhys.91.015005,CHEN202150}. 
The phase with  $\tilde{V} < 1$ is the  topologically non-trivial.
\begin{figure}[htp]
  \begin{centering}
\begin{tabular}{cc}
    \includegraphics[clip,width=0.23\textwidth,angle=0.0]{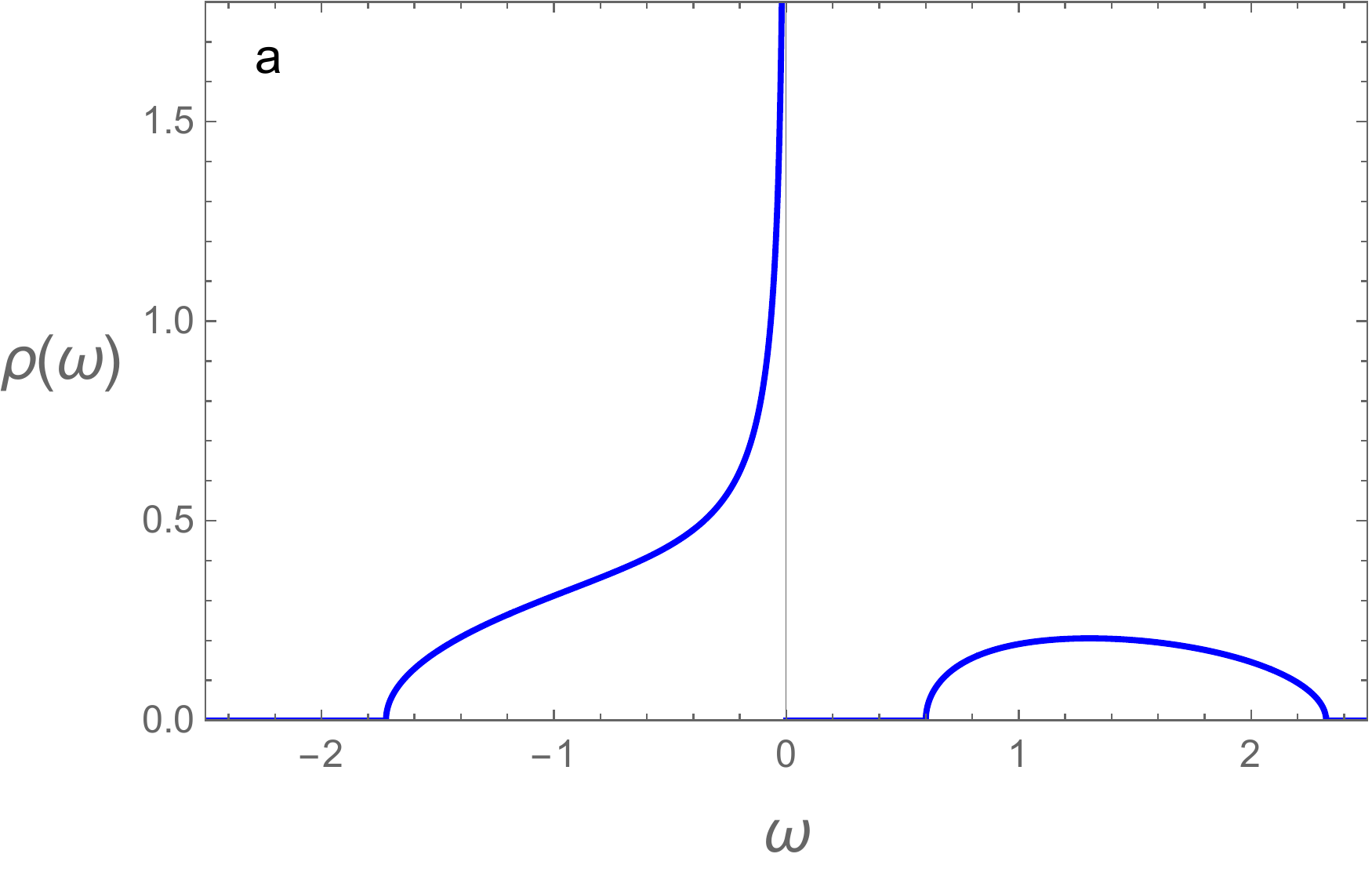} 
     \includegraphics[clip,width=0.23\textwidth,angle=0.0]{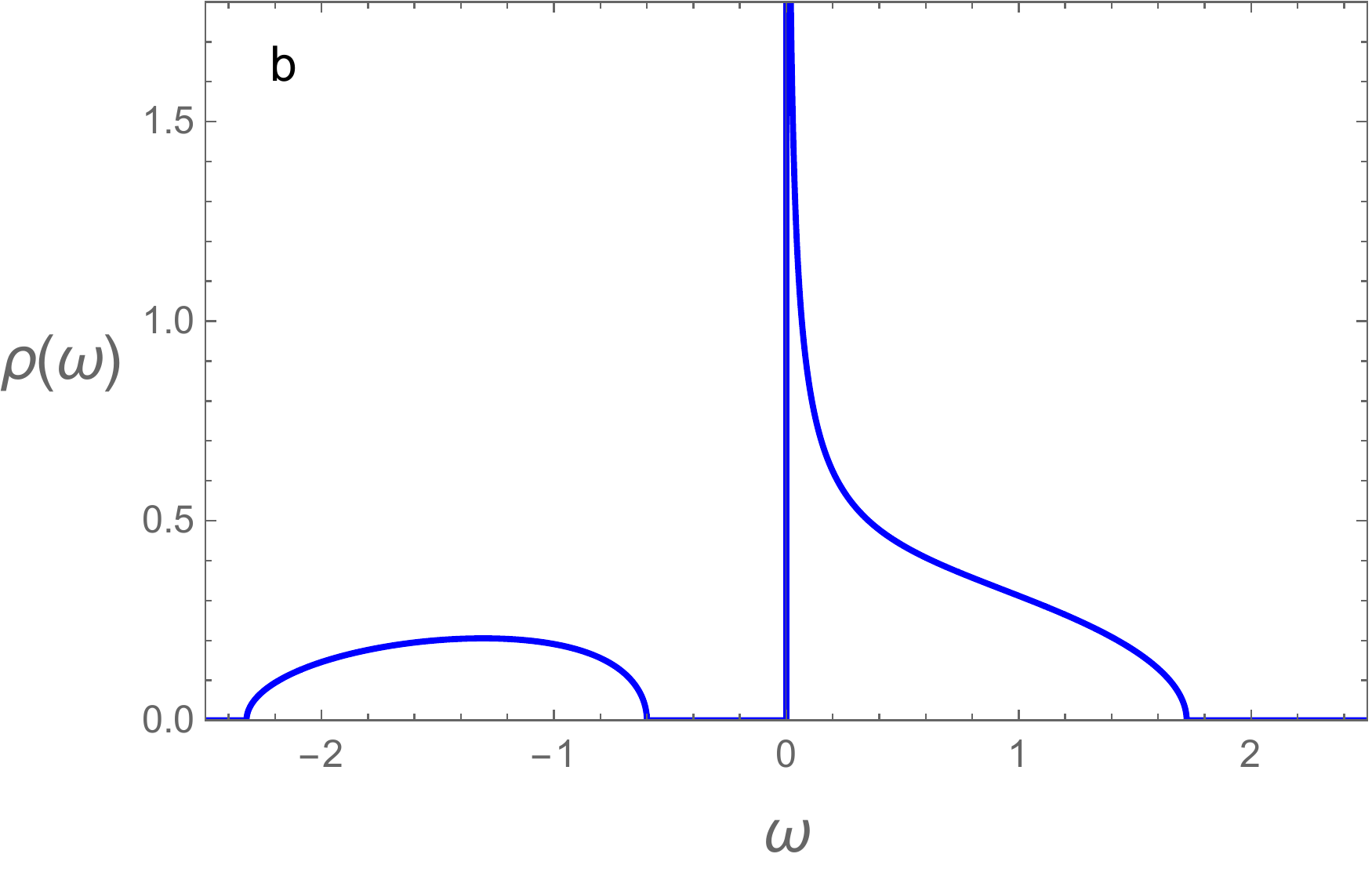}\\ 
     \includegraphics[clip,width=0.23\textwidth,angle=0.0]{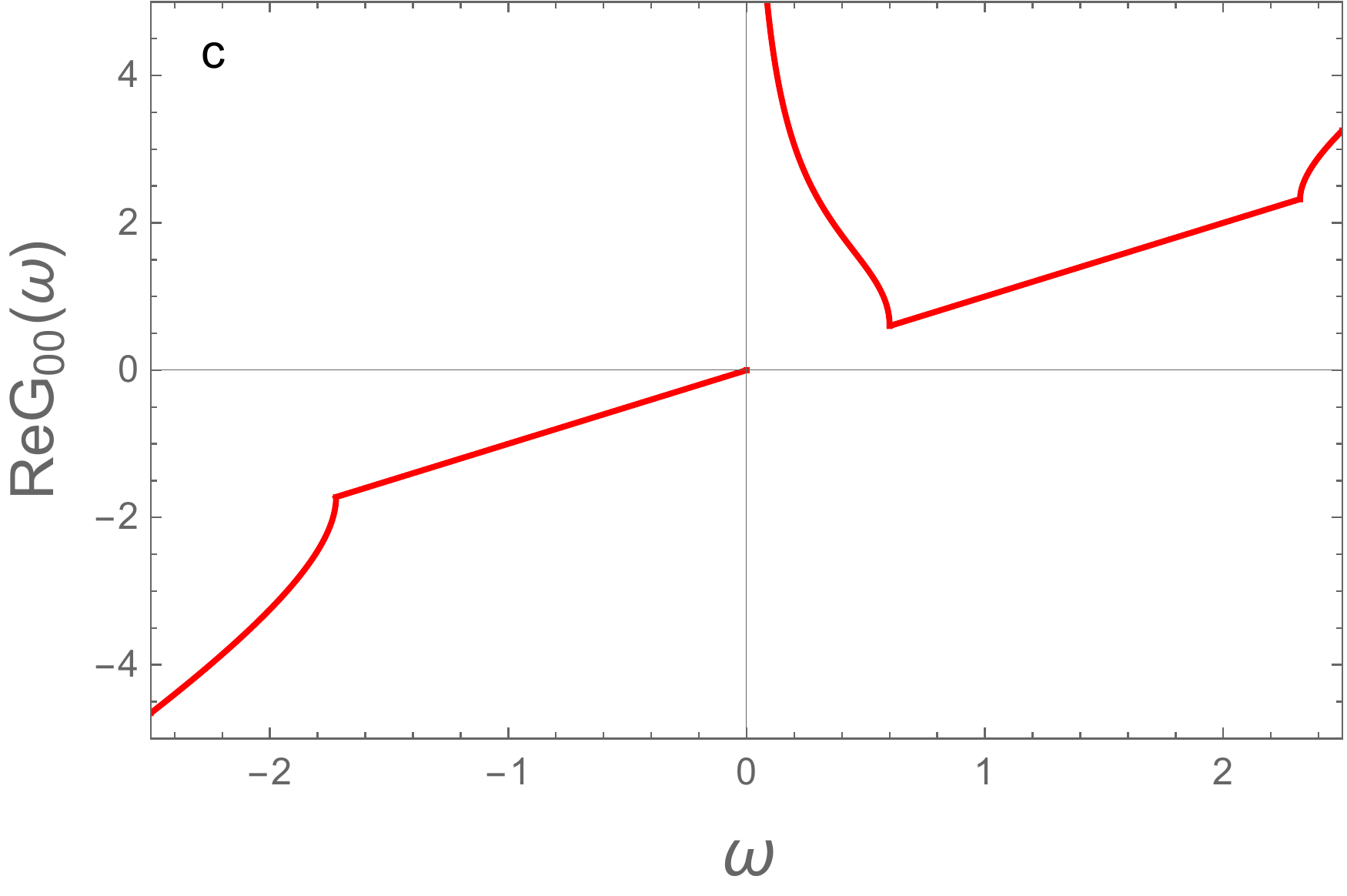} 
   \includegraphics[clip,width=0.23\textwidth,angle=0.0]{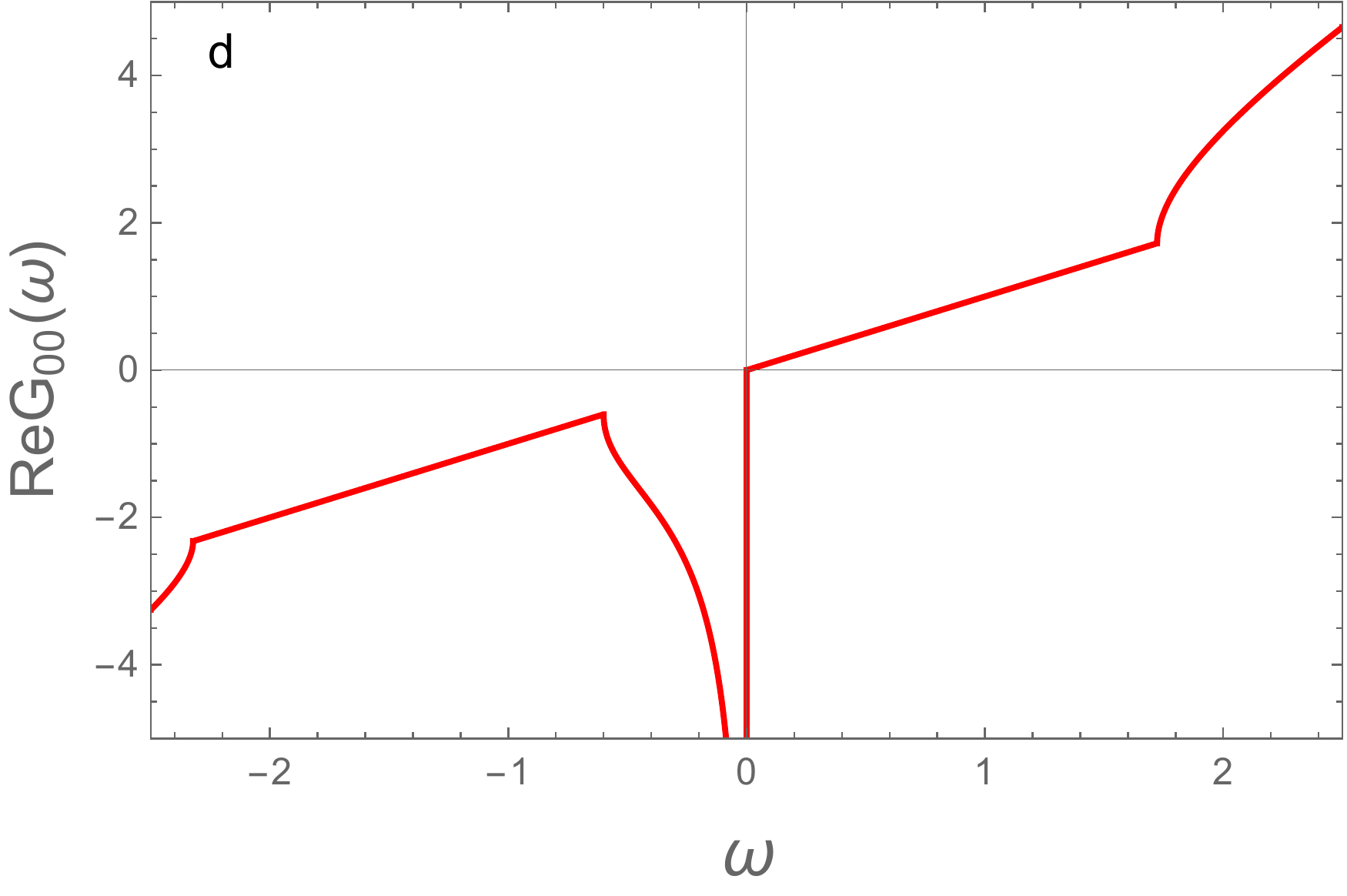} \\
       \end{tabular}
  \end{centering}
    \caption{  \label{fig8n}  (Color online) Surface density of states of the semi-infinite diatomic {\rm sp}-chain (RM chain), Eq.~\ref{rhorm}, at the topological transition ($\tilde{V}=1$), for a) $\mu=\epsilon=- 0.3$ b)  $\mu=\epsilon=0.3$.   Real part of the surface Green's function for c) $\mu=\epsilon=- 0.3$ and d)  $\mu=\epsilon=0.3$. } 
 \end{figure}

Fig~\ref{fig8n} shows the surface density of  states and the real part of the surface Green's function for RM-chains at the topological transition $\tilde{V}=1$. The  figures are for  two values of the energy  of the local surface mode, $\epsilon=\pm 0.1$. The chemical potential is located on the energies of these modes ($\mu=\epsilon$).

\section{Thermoelectric properties of two semi-infinite   chains coupled to a quantum dot} 
\label{sec4}

\begin{figure}[tbh]
\begin{center}
\includegraphics[width=0.5\textwidth,angle=0.0]
{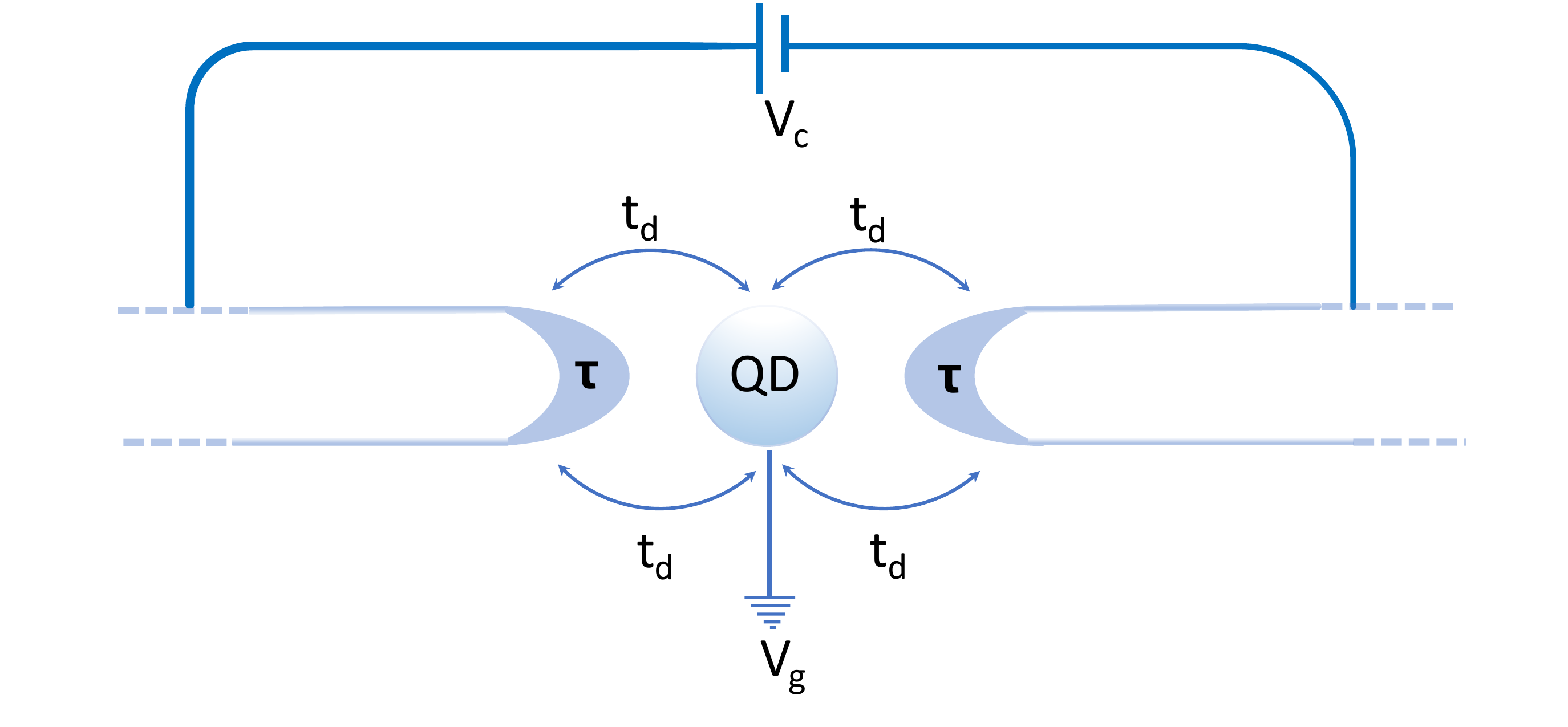}
\end{center}
\caption{\label{qud}(Color online) Two semi-infinite $sp$-chains connected to a quantum dot (QD). A very small potential difference $V_c$  is applied in the chains. Notice that $t_d$ is the coupling between the dot and the chains.}
\end{figure}

In this section we study the transport properties of a device consisting of two identical semi-infinite chains connected to a quantum dot~\cite{maurer2021, PhysRevB.80.235317}, as shown in Fig.~\ref{qud}.  Since we are dealing with spinless fermions, the dot can either be empty, or singly occupied. The non-interacting quantum dot has a single state with energy $E_0$ and is coupled to the chains  by a hopping term $t_d$ that transfers quasi-particles in and out of the dot. Then, the  dot  provides a connection between the semi-infinite chains and   allows  to probe the nature of the edge states through their contribution to the thermal and electrical conductances of the device, as we discuss below.  The coupling Hamiltonian between the dot and the semi-infinite chains is given by, $H_c=-\sum_{\alpha}t_{d,\alpha} c^{+}_{\alpha,0} d+H.c.$, where the  second quantization operators $c$ and $d$ refer to the chains and dot and  $\alpha=r,l$  to the right and left chains, respectively. The dot couples to the first site of each chain (site 0)~\cite{Odashima2017}. For simplicity, we take here  $ t_{d,r}=t_{d,l}=t_{d}$.  

The full local Green's function of the dot connected to the two semi-infinite chains  is given by~\cite{Odashima2017},
\begin{equation}
\label{Gd}
G_d(\omega)=\frac{g_d}{1-2|t_d|^2 g_d G_{00}}
\end{equation}
where,
\begin{equation}
g_d=\frac{1}{\omega-E_0}
\end{equation}
is the Green's function of the non-interacting dot. The Green's function $G_{00}$ is that of the edge of the chains and is given by the self-consistent solution of Eq.~\ref{G000}.
Notice that Eq.~\ref{Gd} can be rewritten as,
\begin{equation}
\label{Gdot}
G_d(\omega)=\frac{1}{\omega -E_0-2|t_d|^2{\rm Re}G_{00}-i2|t_d|^2{\rm Im}G_{00}}.
\end{equation}

The dimensionless electrical conductance of the device, {\it chain-dot-chain} can be obtained as in Ref.~\cite{Ricco2018}. It is given by,
\begin{equation}
\label{sigma}
G/G_0=\int d\omega(-\frac{\partial f}{\partial \omega}) 
\mathcal{T}(\omega)
\end{equation}
where $f(\omega)$ is the Fermi-Dirac distribution and
$$\mathcal{T}(\omega) =-\Gamma {\rm Im} G_d(\omega)$$
with 
$\Gamma= 2 \pi |t_d|^2  {\rm Im} G_{00}$, the {\it Anderson broadening}~\cite{Ricco2018}.
In Eq.~\ref{sigma}, $G_0=e^2/h$ is the quantum of conductance.

More generally, we  define the quantities,
\begin{equation}
\label{Ln}
\mathcal{L}_n=\frac{1}{h} \int d\omega(-\frac{\partial f}{\partial \omega}) \omega^n
\mathcal{T}(\omega),
\end{equation}
in terms of which we can obtain the thermoelectric coefficients. The conductance can be rewritten as
$G=e^2\mathcal{L}_0$.
The thermal conductance $K$ and the thermopower  $S$ are  given, respectively,  by
\begin{equation}
\label{KT}
K=\frac{1}{T} \left( \mathcal{L}_2 - \frac{\mathcal{L}_1^2} {\mathcal{L}_0}  \right),  
\end{equation}
\begin{equation}
\label{TP}
S=-\left( \frac{1}{e T}\right)
  \frac{\mathcal{L}_1}{\mathcal{L}_0}. 
  \end{equation}
These in turn define the   Wiedemann-Franz ratio (WF)  and the dimensionless {\it figure of merit}  $ZT$ that are given, respectively, by
\begin{equation}
\label{wf}
WF=\frac{1}{T} \left(\frac{K}{G}\right),
\end{equation}
\begin{equation}
\label{zt}
ZT=\frac{S^2GT}{K},
\end{equation}
where the  former ratio WF is given in units of the Lorenz number $L_0=(\pi^2/3)(k_B/e)^2$.

The Mahan-Sofo parameter  $\zeta$ \cite{Sofo} is defined in terms of the  thermoelectric coefficients 
\begin{equation}
\zeta = \frac{L_{1}^{2}(T)}{L_{0}(T)L_{2}(T)} ,
\label{Factor}
\end{equation}
and using this parameter,  the  dimensionless thermoelectric figure of merit, defined 
in Eq. (\ref{zt}), can be written as
\begin{equation}
ZT=\frac{\zeta}{1-\zeta}.
\label{ZT10}
\end{equation}
The best $ZT$ occurs at the limit $\zeta \rightarrow 1$.

\section{Results  for SSH chains or monoatomic sp-chains}
\label{sec5}

We start obtaining the thermoelectric properties of the device in the case the dot is coupled to SSH chains. We calculate, using the equations above,  the thermoelectric properties  of the coupled system, dot-chains, in the different topological phases of the SSH chains and at the topological transition.
When the chains are in either the trivial or topological phases,  i.e., for   $\tilde{V}>1$ and  $\tilde{V}<1$,  respectively,  the conductances  are zero at zero temperature, since the bulk of the chains are insulators. At finite temperatures these conductances become finite due to thermal activation of  quasi-particles above the band gap. 
The results presented  are obtained for the chemical potential of the chains $\mu=0$, i.e., for a full lower band ({\it half-filling}). The dot energy is  $E_0=0$, and the coupling between the dot and the chains is taken as, $t_d/V_2=0.15$.

\begin{figure*}[ht]
\begin{center}
\includegraphics[width=1\textwidth,angle=0.0]
{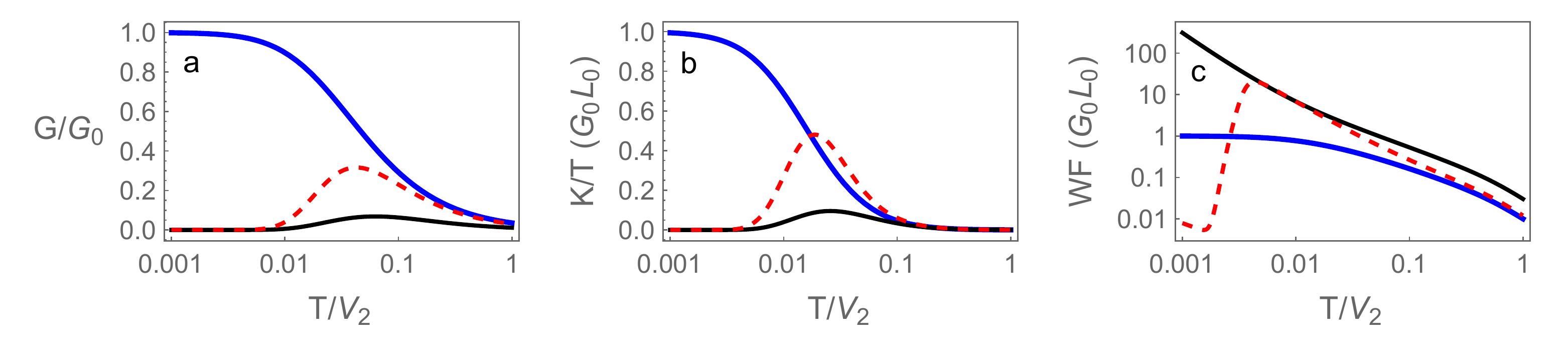}
\end{center}
\caption{ \label{fig4} (Color online)  a) Dimensionless electrical conductance, b) thermal conductivity divided by temperature in units of $G_0L_0$ and c) Wiedemann-Franz ratio  ($WF=(K/T)/(G/G_0)$)  in units of $G_0L_0$ as functions of temperature for the device with SSH chains.  In the trivial phase ($\tilde{V}=1.03$) black continuous, topological phase ($\tilde{V}=0.97$) red dashed   and at the topological transition ($\tilde{V}=1$) blue continuous.}
\end{figure*}

Fig.~\ref{fig4}a shows the conductance of the device in the trivial and topological phases. As expected the conductances vanish at $T=0$ in both phases and become finite at finite temperatures. The finite temperature conductance is larger in the topological phase. Notice that in both cases shown, $\tilde{V}=1.03$ and $\tilde{V}=0.97$, the system is at the same distance of the topological transition at $\tilde{V}=1$. The   increment of the conductance in the topological phase can be attributed to the presence of the edge mode.
 
Fig.~\ref{fig4}b  shows the thermal conductivity divided by temperature, in units of $G_0L_0$. They also vanish for $T=0$, in both the trivial and topological phases,  as expected, since the {\it bulk} of the chains is insulating.  

Fig.~\ref{fig4}c  shows the Wiedemann-Franz ratio, defined as $WF=(K/T)/(G/G_0)$ and  in units of $G_0L_0$, as a function of temperature. Away from the topological transition in both trivial and topological phases the Wiedemann-Franz law is violated. This can occur in topological systems~\cite{buccheri2021violation, giuliano2021multiparticle}, as for the monoatomic chains  and in general for diatomic chains, as   we discuss below and show in Fig.~\ref{Activated}.

Fig.~\ref{fig4}a, b  and c, also show the conductance, thermal conductivity and Wiedemann-Franz ratio {\it at the topological transition}, i.e., at $\tilde{V}=1$.  The zero temperature dimensionless electrical conductance in this case is unity showing that a quantum of charge flows through the system. Then, at the transition the surface modes recombine to form a quasi-particle  that  transports electric current through the dot. The current can flow through the device since, at $V_1=1$, 
the chains are in a semi-metallic state (Dirac semi-metal).   We point out that the zero temperature electrical conductance at the transition  does not depend  on the coupling  $t_d$ between the dot and the chains.
The thermal conductance $K$, differently from the electrical conductance vanishes at zero temperature, even at the topological transition. However,  the temperature normalized thermal conductance ($K/T$) at the topological transition goes in this limit to $1$, in units of $G_0L_0$ ,  as shown in Fig.~\ref{fig4} b.  
The Wiedemann-Franz ratio, at the topological transition of the monoatomic chain,   starts as unity at $T=0$ and remains constant at very low temperatures showing that the Wiedemann-Franz law is obeyed in this case.

Finally, we remark that the thermopower, Eq.~\ref{TP}, vanishes  at the trivial and topological phases and also  at the topological transition. This occurs since the quantity $\mathcal{L}_1$ in this equation cancels out due to equal but opposite contributions of electrons and holes to this quantity in this particle-hole symmetric case.
 
\section{Results of the diatomic  $sp$ or Rice-Mele chains }
\label{sec6}

\begin{figure}[tbh]
\begin{center}
\includegraphics[width=0.45\textwidth,angle=0.0]
{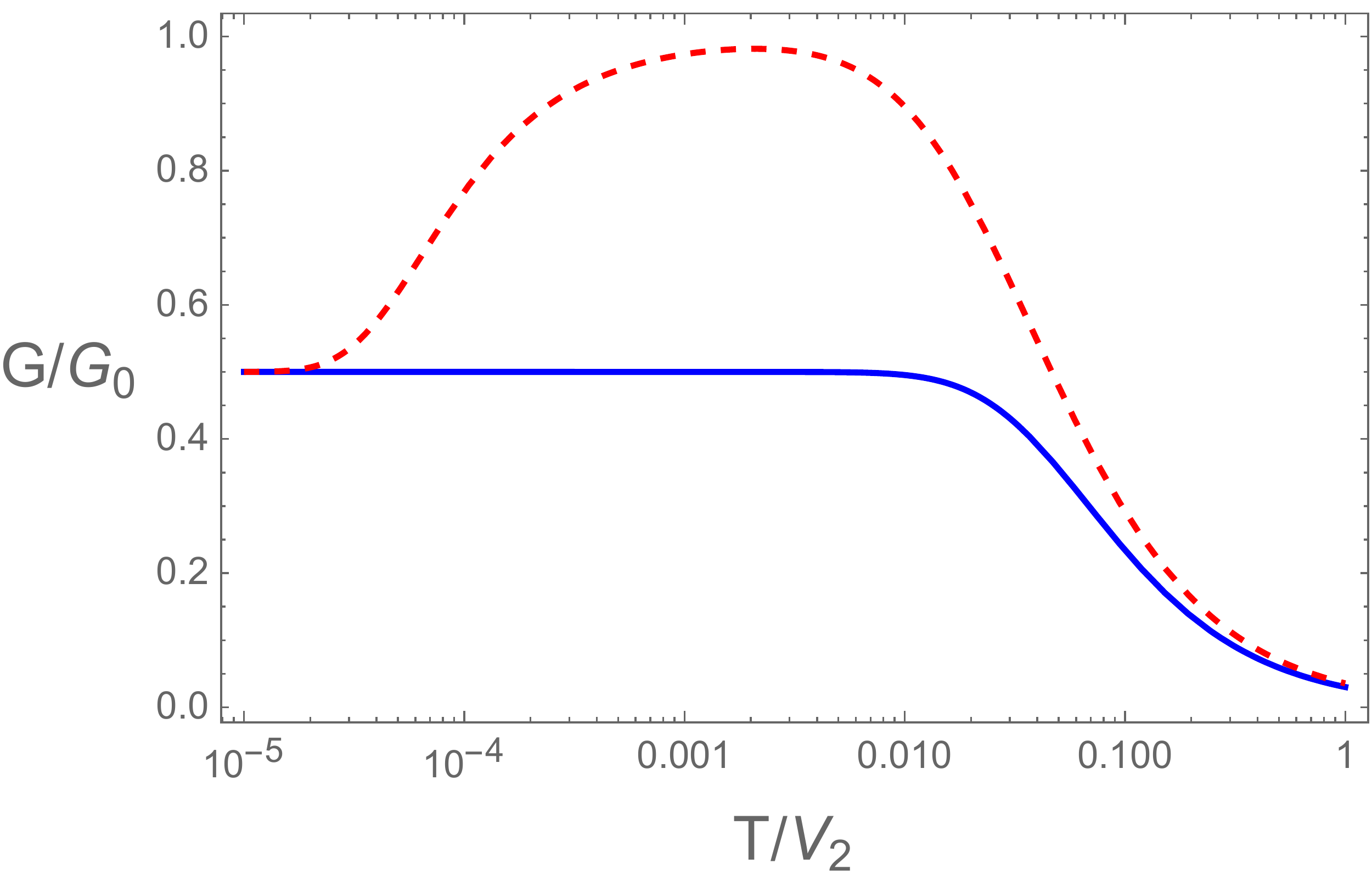}
\end{center}
\caption{\label{figGG00}(Color online) Normalized conductances as a function of temperature for a system consisting of two semi-infinite Rice-Mele chains attached to a quantum dot at the topological phase transition of the chains ($\tilde{V}=1$).  In blue continuous $\epsilon/V_2=0.1$, and in red dashed  $\epsilon/V_2=5 \times 10^{-5}$. In both cases, the low temperature saturation value $G/G_0=1/2$    gives evidence of fractional charges  flowing in the system. The curves for $G/G_0$ are independent of the coupling to the quantum dot and of the sign of $\epsilon$, for $\mu=\epsilon$ and the dot in resonance with the edge mode. Notice that for small values of $\epsilon/V_2$ as  temperatures increases there is a kind of recombination of the fractional charges. }
\end{figure}

In this section we obtain the thermoelectric properties of the device when Rice-Mele chains are attached to the quantum dot. Notice that in this case the chiral symmetry of the SSH chain is broken for RM chains. We consider the situation where the chemical potential coincides with the local energy of one of the sub-lattices, i.e., we take $\mu=\pm \epsilon$.  Furthermore we consider that the quantum dot is in resonance with the energy of the edge mode, which for the condition  $\mu=\pm \epsilon$ corresponds to take $E_0=0$. Since $\epsilon \ne 0$, the topological transition of the model occurs for $\tilde{V}=1$. We start  showing the 
normalized temperature dependent conductances   of the Rice-Mele model at the topological transition. As can be seen in Fig.~\ref{figGG00}, the normalized conductances at zero temperature now attains a value of $1/2$, expected when  fractional charges $e/2$ are responsible for the electronic transport in the system. This result is universal in the sense that it is independent of the coupling $t_d$ between the dot and the chains and the value of $\epsilon$, for the conditions specified above ($\mu=\pm \epsilon$, $E_0=0$). Whenever we use this term here we refer to this type of universality. The figure shows the normalized conductance for two values of $\epsilon/V_2$. Notice that for $\epsilon/V_2  \ll 1$ the finite temperature  conductance reaches a maximum value close to one, as if there is a recombination of the fractional charges in the system due to thermal effects. 

We point out that the  fractional charge as evidenced by the zero temperature conductance is a direct consequence of the breaking of chiral symmetry  of the original SSH model, due to the finite and distinct energies of the sub-lattices of the Rice-Mele model.

\subsubsection{Thermopower}

\begin{figure}[tbh]
\begin{center}
\includegraphics[width=0.45\textwidth,angle=0.0]
{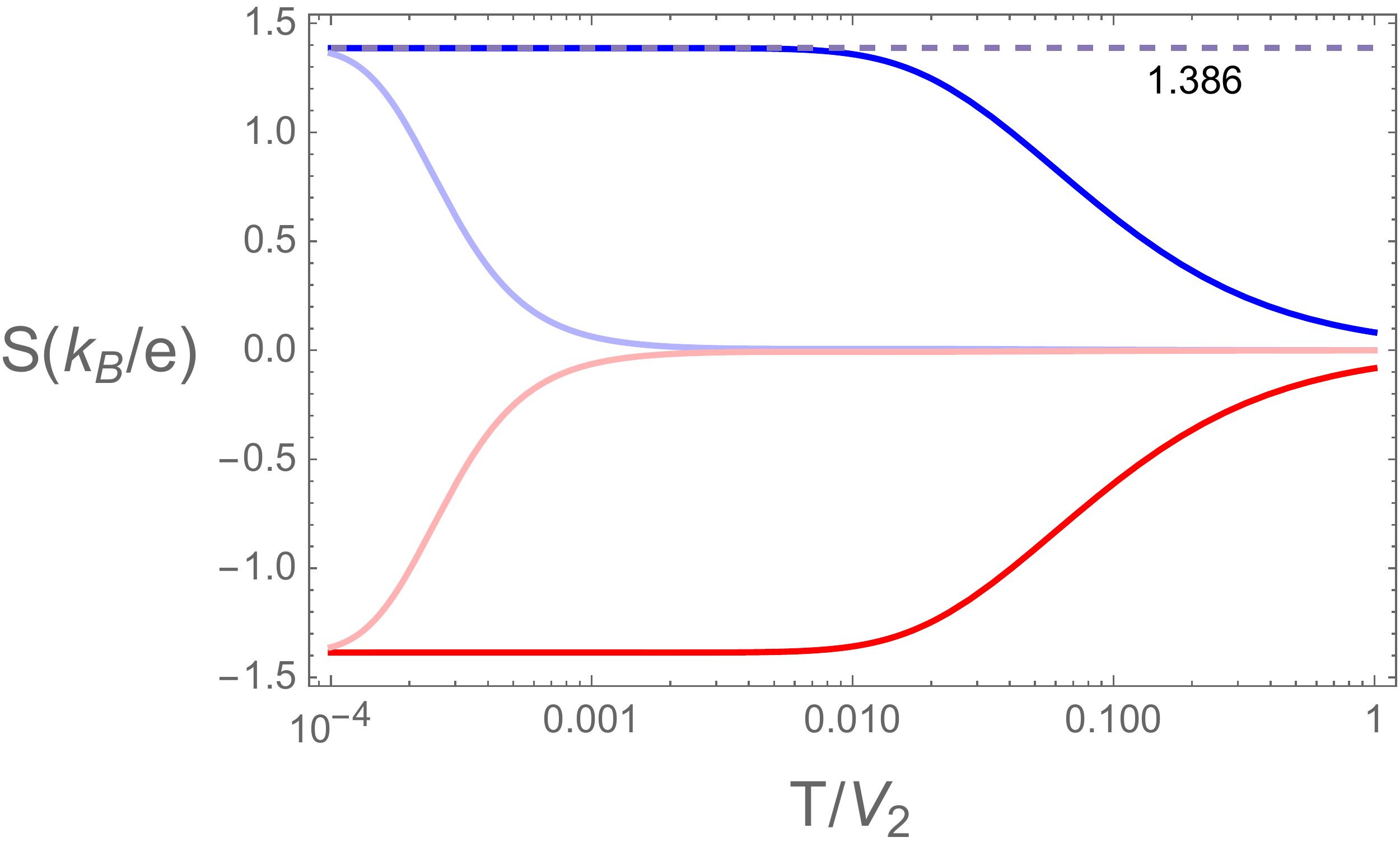}
\end{center}
\caption{\label{figTP} (Color online) Thermopower of the device as a function of temperature in units of $(k_B/e)$ at the topological  transition of the RM chains. $S>0$ correspond to $\mu=\epsilon=+0.1$, and $\mu=\epsilon=+3.3 \times 10^{-4}$  (light curve). Negative thermopower ($S<0$) corresponds to   $\mu=\epsilon=-0.1$ and  $\mu=\epsilon=-3.3 \times 10^{-4}$  (light curve).  The light color curves show the trend to the results the SSH chain with $\epsilon=0$. The energy scale for the low temperature saturation of the thermopower is given by the difference   in site energies, $2 \epsilon$. The numerical results for the saturation values, $S(T=0) \approx  \pm 1.38634$  are in close agreement with $S_0=\pm 2 \ln 2 \approx \pm 1.38634$,  as discussed in the text.  } 
\end{figure}

The thermopower is an  interesting and unique physical property that contains fundamental information on both,  transport and thermodynamic properties of the system. The temperature dependence of the thermopower of the device consisting of two Rice-Mele chains coupled to the quantum dot can be obtained using Eq.~\ref{TP}.  At  the topological transition ($\tilde{V}=1$),  this  is shown in Fig.~\ref{figTP} for $\mu = \epsilon$ and the cases of $\epsilon$ positive and negative. The corresponding surface density of states for these two cases is shown in the upper panels of Fig.~\ref{fig8n}.  The thermopower is  positive or negative depending whether the charge carriers are holes or electrons, respectively. It is constant at low temperatures and its absolute value decreases with increasing temperature. It is remarkable that it does not vanish for  $T \rightarrow 0$, as expected from the third law of thermodynamics.  Mathematically, this arises since the function $\mathcal{T}(\omega) $ in Eq.~\ref{Ln} has a jump  discontinuity  and is non-differentiable at $\omega =0$, which precludes a low temperature Sommerfeld expansion.

The constant low temperature values for the thermopower, $S(T \rightarrow 0) \approx \pm 1.386$ can be rationalized in terms of the properties of the quantum dot and of the chains at the topological transition. Since we took $\mu=\epsilon$, the doubly degenerate zero energy surface mode~\cite{Kempkes:2016ww} becomes delocalized at the transition and every site in the system including the dot has a double degenerescence. 
For a  system of charged particles, the thermopower represents the entropy per carrier divided by the charge of the carrier~\cite{Chaikin1990},
\begin{equation}
\label{show}
S_0=\frac{\text{entropy per carrier}}{q^*}.
\end{equation}
This is also known as the Kelvin formula for the Seebeck coefficient~\cite{Shastry_2010}. The entropy per site is $S= \ln 2$ and remains finite at $T=0$ due to the double degeneracy of the states,  whether a site is occupied by a particle or by a hole.    If the carriers have a fractional charge, $q^*= \pm 1/2$ (in units of electric charge) as evidenced by the zero temperature conductance,  we get 
\begin{equation}
\label{S0}
S_0=\frac{\ln 2}{(\pm 1/2)}=\pm 2 \ln 2 \approx \pm 1.386 (k_B/e),
\end{equation} 
which are exactly the low temperature saturation values, obtained   numerically for the thermopower using Eq.~\ref{TP},  as shown in Fig.~\ref{figTP}. These values are universal in the same sense we used for the conductance, i.e., they are independent of  $\epsilon$ and $t_d$ (for $\mu=\epsilon$, $E_0=0$). 
\begin{figure}[tbh]
\begin{center}
\includegraphics[width=0.45\textwidth,angle=0.0]
{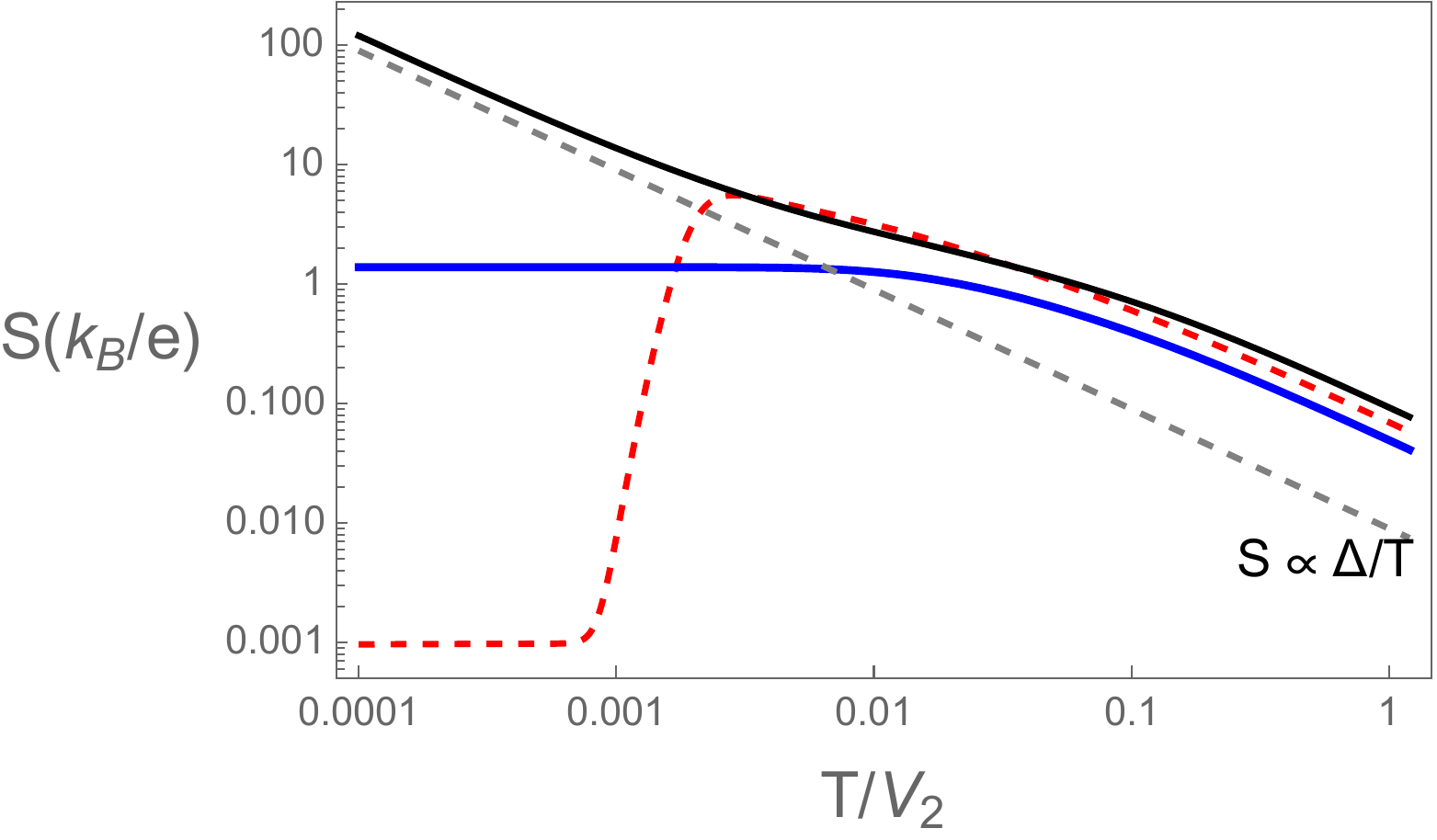}
\end{center}
\caption{\label{Snew3} (Color online) Thermopower of the device as a function of temperature in units of $(k_B/e)$ away and at  the  topological  transition of the RM chains. Red dashed corresponds to $\tilde{V}=0.95$, such that the chains are in the topological phase. Black continuous  shows the thermopower in the trivial phase, with $\tilde{V}=1.05$ and blue at the topological transition. The gray dashed line shows the classical result for a semiconductor with activation energy $\Delta$. } 
\end{figure}

Then this result for the thermopower, together with that for the electrical conductance, corroborate the existence of  carriers with fractional charges, $q=\pm e/2$, flowing in the device with RM chains at the topological transition.   This  {\it transport charge} does not necessarily coincide with the concept of  {\it boundary charge}~\cite{PhysRevB.102.085122}.
This is clear, since at the topological phase transition where our results are obtained,  the penetration depths of the edge modes diverge and their charge is spread all over the system~\cite{Continentino2014c, PhysRevB.100.195432, sabrina2021}.

For completeness we show in Fig.~\ref{Snew3} the temperature dependent thermopower away from the topological transition in both trivial and non-trivial topological phases.

\subsubsection{\label{GTWF}Thermal conductance and Wiedemann-Franz ratio}

The thermal conductance divided by temperature (K/T) at the topological transition of the diatomic $sp$-chain is shown in Fig.~\ref{figmucio4}. From Eqs.~\ref{Ln} to~\ref{wf}, we can write
\begin{equation}
\frac{K}{T} = \frac{\mathcal{L}_2}{T^2}-  \left( \frac{\mathcal{L}_1}{e T \mathcal{L}_0}\right)^2  e^2 \mathcal{L}_0,
\end{equation}
and using the expressions for the thermopower and conductance we get,
\begin{equation}
\label{KTRM}
\frac{K}{T} = \frac{\mathcal{L}_2}{T^2}-  S^2 G.
\end{equation}
\begin{figure}[ht]
  \begin{centering}
    \includegraphics[clip,width=0.42\textwidth,angle=0.0]{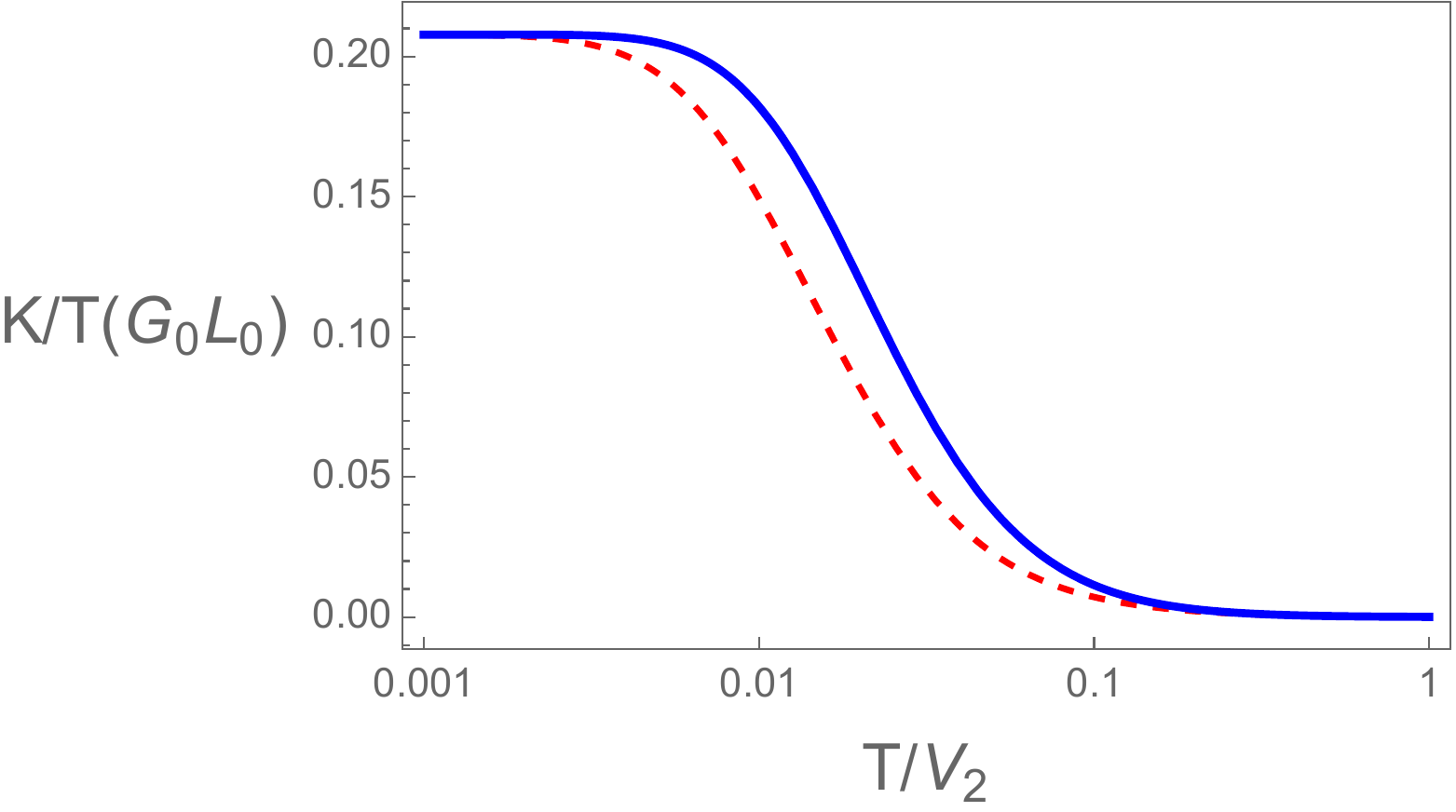} 
    \end{centering}
     \caption{ \label{figKTRM}  (Color online) Thermal conductance divided by temperature of the device at the topological transition of the RM chains in units of $G_0L_0$,  where $L_0$ is the Lorenz number for $\epsilon=01$ blue continuous, and $\epsilon=0.3$ red dashed.  The zero temperature limiting value $(K/T)_0 \approx 0.20792$ (see text). is independent of the values of $t_d$ and $\epsilon$ as long as, $\mu=\epsilon$ and $E_0=0$.  }
 \end{figure}
We can obtain  the limit of zero temperature analytically, $(K/T)_0 = ( K/T)_{T \rightarrow 0}$, using the results for the thermopower, Eq.~\ref{S0}, and for the conductance. We find,
\begin{equation}
(K/T)_0=   \frac{1}{2}(1 - \frac{3}{\pi^2}(2 \ln 2)^2) \approx 0.20792,
\end{equation}
in units of $G_0L_0$. This is in agreement with the numerical result shown in Fig.~\ref{figKTRM} and it is independent of $\epsilon$ and $t_d$.
The dimensionless Wiedemann-Franz ratio  attains at zero temperature the value, $\mathcal{W}= (WF/L_0)= 1/2$.  This value of $\mathcal{W}$  is different from that for metallic chains where $\mathcal{W}=1$.  Violation of the Wiedemann-Franz law has been found in interacting systems~\cite{buccheri2021violation,ma14112704} and in devices with interacting quantum dots~\cite{Majek2022,kubala2008}. 

For completeness, we point out that away from the topological transition, both in the trivial and non-trivial topological phases we obtain that the conductance and thermal conductivities are thermally activated as in a semi-conductor.

\subsection{Figure of merit and power factor}

\begin{figure}[ht]
\begin{center}
\includegraphics[width=0.49\textwidth,angle=0.0]
{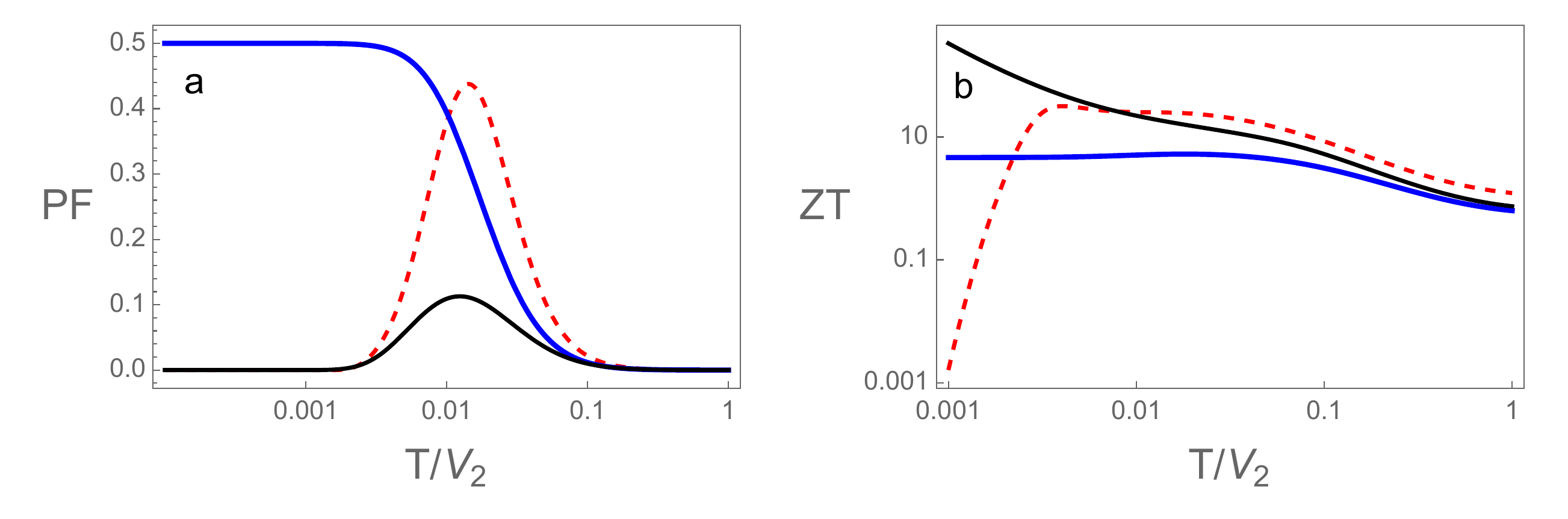}
\end{center}
\caption{ \label{figmucio4} (Color online)  a) Dimensionless electrical conductance, b) thermal conductivity divided by temperature in units of $G_0L_0$ and c) Wiedemann-Franz ratio  ($WF=(K/T)/(G/G_0)$)  in units of $G_0L_0$ as functions of temperature for the device with SSH chains.  In the trivial phase ($\tilde{V}=1.03$) black continuous, topological phase ($\tilde{V}=0.97$) red dashed   and at the topological transition ($\tilde{V}=1$) blue continuous.}
\end{figure}

Fig.~\ref{figmucio4}   shows the dimensionless  power factors~\cite{mahan1996} and figures of merit $ZT=(S^2GT)/K$   of the device,  as functions of temperature, at the topological transition, $\tilde{V}=1$, and in the trivial $\tilde{V}=1.05$ and topological $\tilde{V}=0.95$ phases of the RM chains. 
The power factor is defined as $PF=(\widetilde{PF}/S_0^2 G_0)$, where $S_0$ is the zero temperature thermopower and $G_0$ the unit of conductance. The quantity $\widetilde{PF}=S^2G$ is the full dimensional power factor~ \cite{Benenti2017}.  These quantities ZT and PF do not depend on the  sign of $\epsilon$,  only on its absolute value. 

Notice that the figures of merit ZT at the trivial and topological phases assume large values, for the parameters used in Fig.~\ref{figmucio4} at a temperature of$T/V_2 \approx 0.01$, where the power factor is close to a maximum. In order to translate this in physical temperature notice that the energy scale $V_2$ is of the order of a bandwidth ($ \sim 1$ eV or $\sim 10^4$ K). In the trivial semiconductor phase although the figure merit increases at lower temperatures, the power factor drops to very small values, while it continues significant at the topological transition.    
The significance of this quantity (PF)  is that, in a time reversible system at steady state, the maximum power for conversion of heat into work is given by $P_{max}= (1/4) \widetilde{PF}$ for two heat reservoirs with a difference in temperature $\Delta T=1$ K. The efficiency of a device at this maximum power is given by~\cite{Benenti2017}, 
\begin{equation}
\eta(P_{max})=\frac{\eta_{ca}}{2} \frac{ZT}{ZT+2}
\end{equation}
where $\eta_{ca}$ is the efficiency of a Carnot engine working between the same reservoirs.

It is worth emphasizing that the relevant characteristic temperatures we obtain, for example,  for the saturation of the thermopower at low temperatures,  maxima of PF, saturation of WF using reasonable values for the parameters of the dot-chains system  are much larger than the actual Kondo temperature of realistic quantum dots~\cite{svilans2018,dutta2019}.

\section{High temperature results}

\label{sec7n}

In this section we present the results for the thermally activated thermoelectric properties of the device coupled to RM chains.

\begin{figure}[tbh]
\begin{center}
\includegraphics[width=0.45\textwidth,angle=0.0]{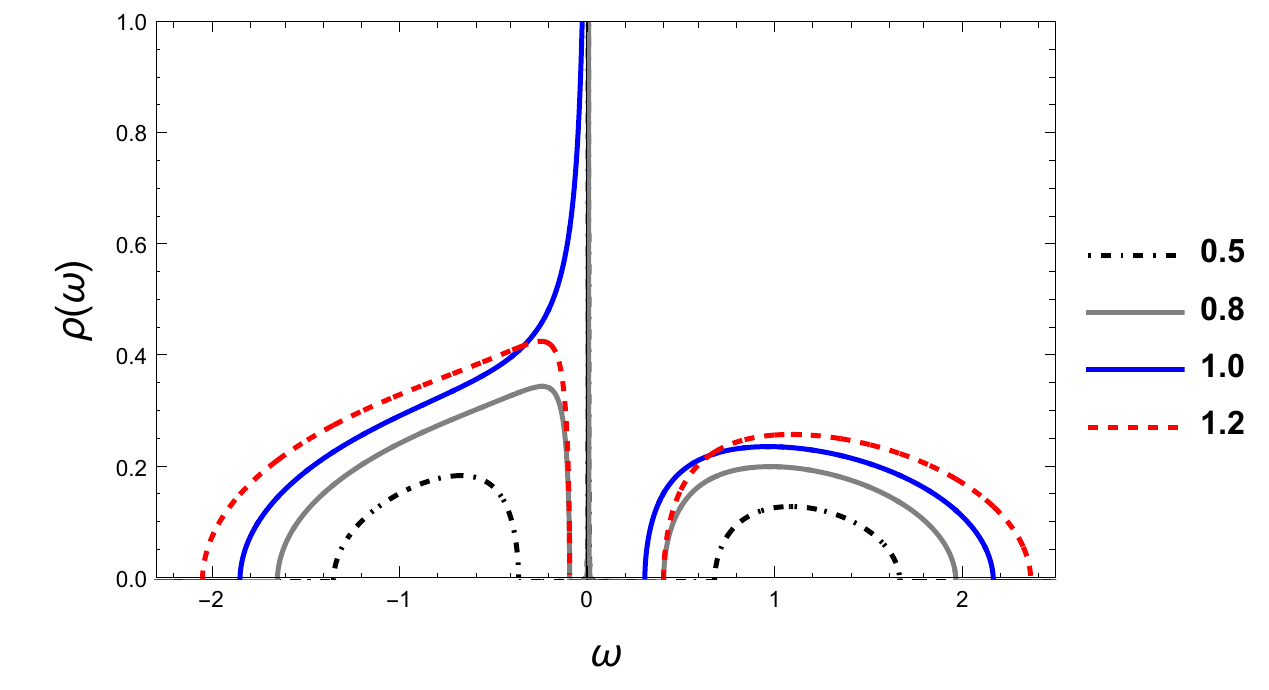}
\end{center}
    \caption{(Color online)  Density of states corresponding to different values of $V_{1}/V_{2}=1.2;1.0;0.8;0.5$. The legends represent the values of $\tilde{V}=V_1/V_2$.}
\label{Gap}
 \end{figure}
In Fig.\ref{Gap} we present the density of states for different values of the ratio $V_{1}/V_{2}=1.2;1.0;0.8;0.5$.  Two points should be noticed here: First, at the topological phase transition,  $V_{1}/V_{2}=1.0$, the density of states (red curve) presents a sharp behavior at $\mu=0$ that gives rise to an electrical conductance $G/G_{0}=0.5$ at low temperatures. On the contrary, inside the topological region, the density of states at the chemical potential presents a delta function, as indicated in the curves with $V_{1}/V_{2}=0.8,0.5$. On the other hand, the curve $V_{1}/V_{2}=1.2$, outside the topological region, exhibits a full gap. The second point, and the most important for our purposes, is that inside the topological region, as $V_{1}/V_{2}$ decreases, the electrons migrate from the valence band to the peak located at the chemical potential, increasing its weight and the gap, allowing for tuning the thermoelectric properties to the room temperatures region.

\begin{figure}[htb]
\centering
  \begin{tabular}{@{}cccc@{}}
  \includegraphics[width=.23\textwidth]{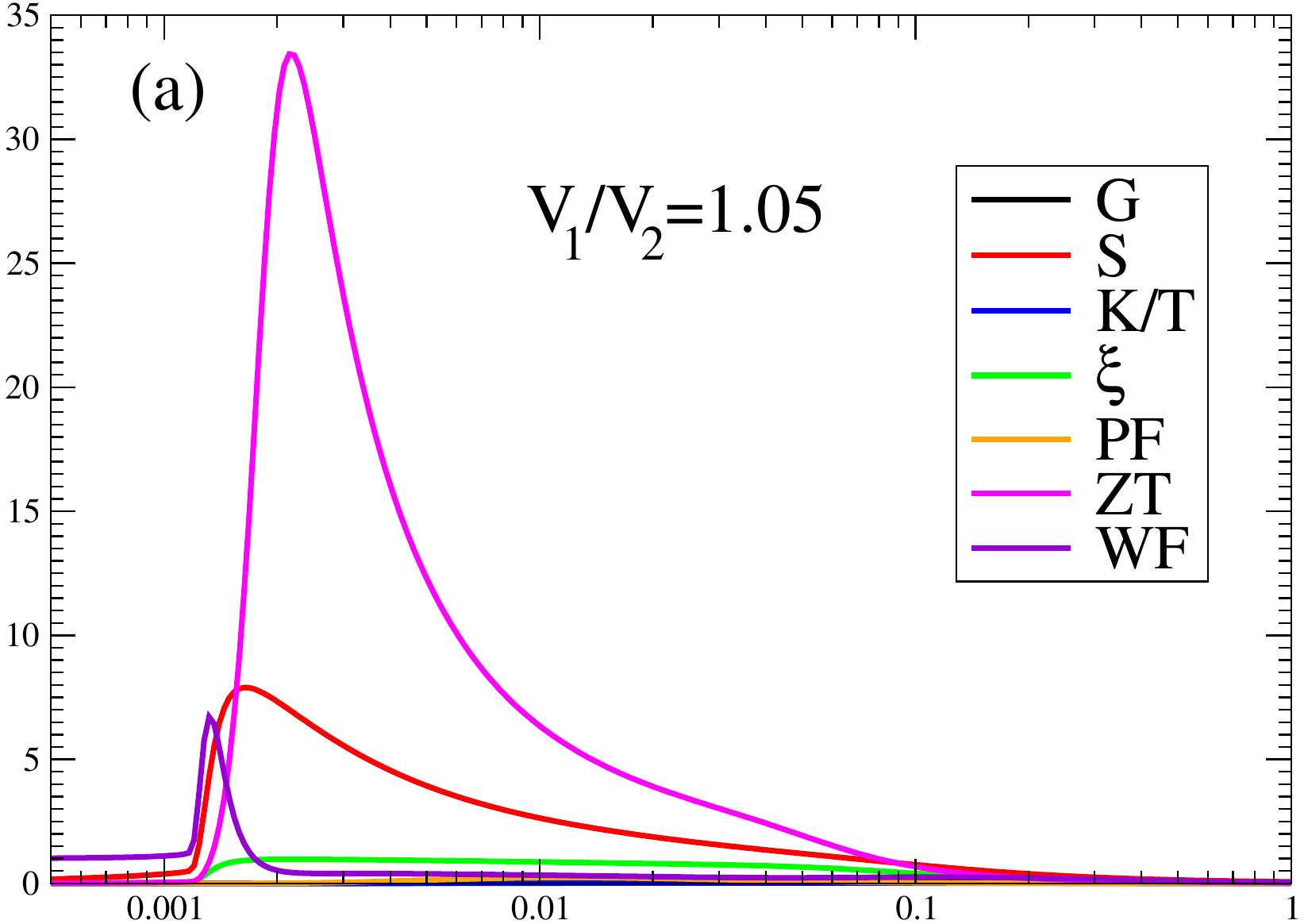} &\
  \includegraphics[width=.23\textwidth]{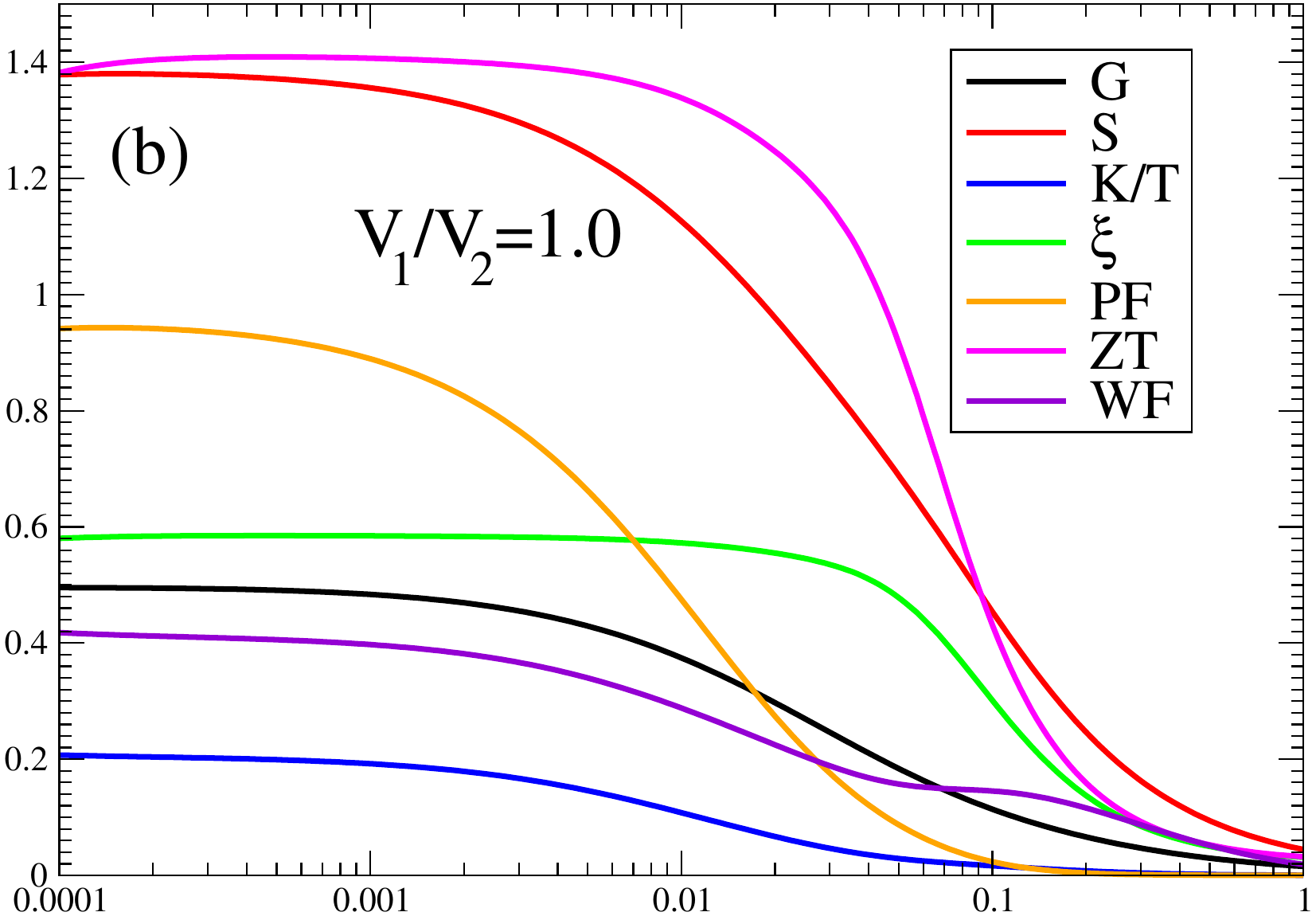}&\\ 
  \includegraphics[width=.23\textwidth]{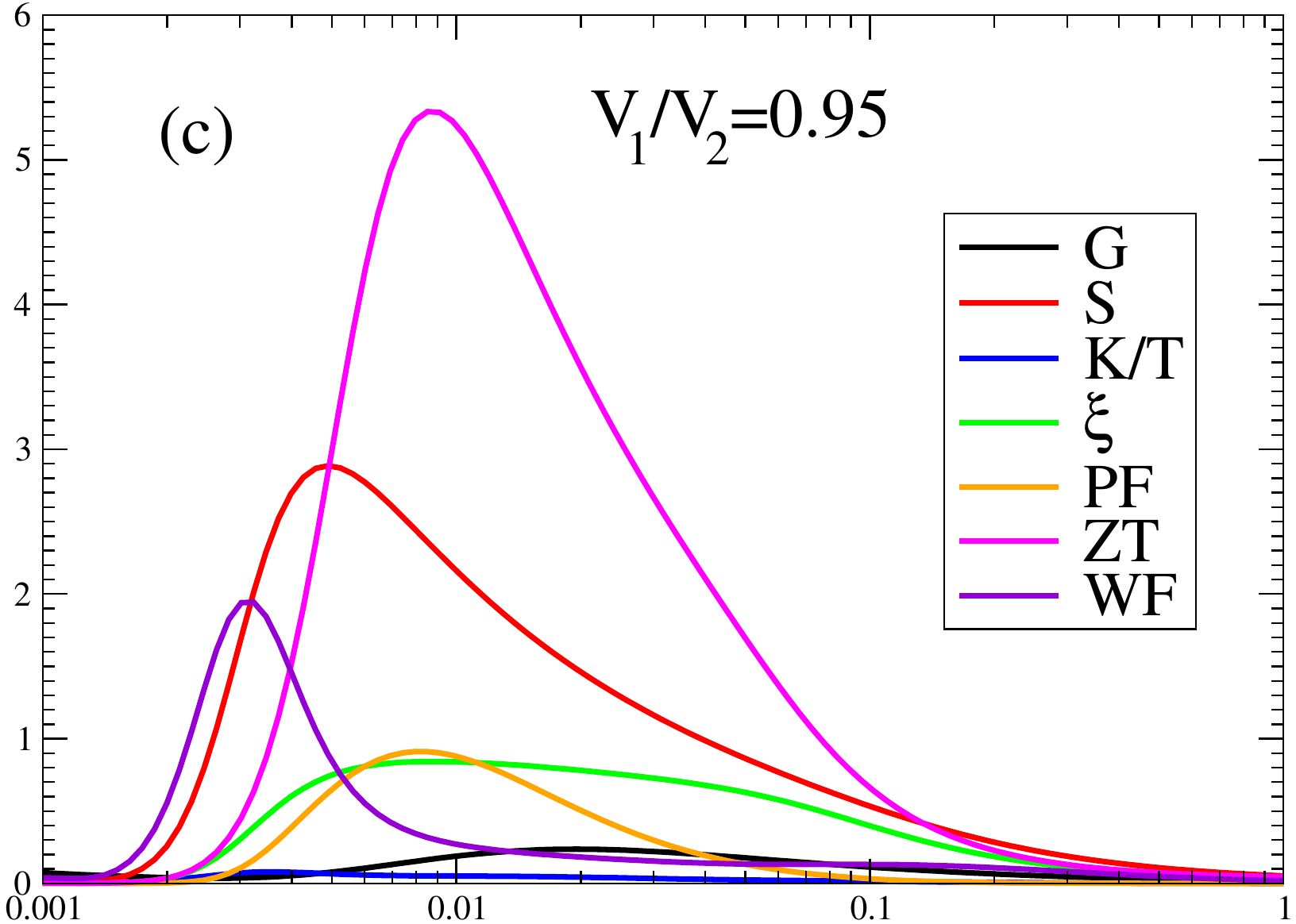} &
  \includegraphics[width=.23\textwidth]{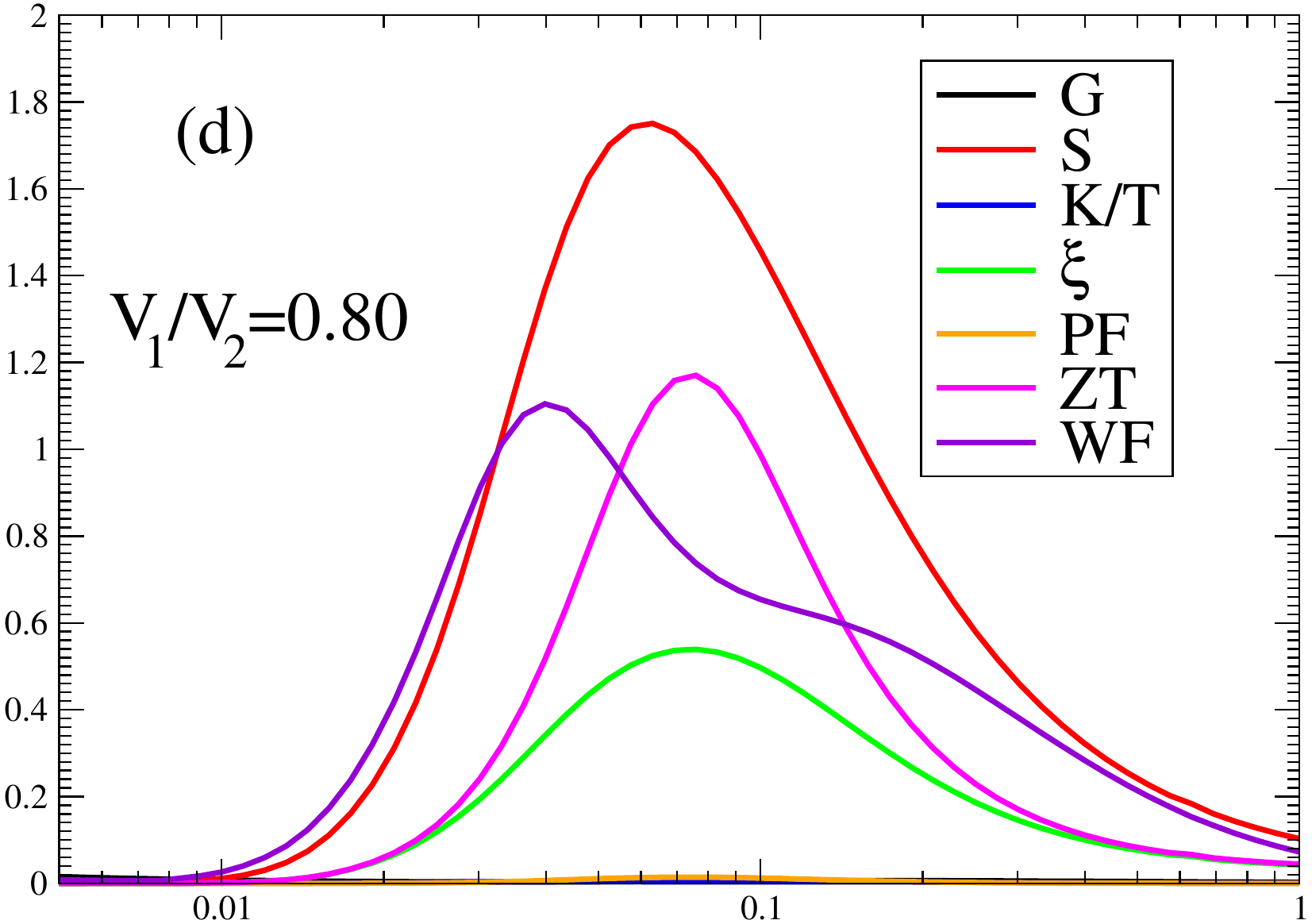}   \\
 \end{tabular}
  \caption{(Color online)  Temperature activated thermoelectric properties for different values of the hybridization $V_{1}/V_{2}: a)1.2; \hspace{0.1cm}  b)1.0;  \hspace{0.1cm} c) 0.8;  \hspace{0.1cm} d) 0.5$. }
\label{Activated}
\end{figure}

In Figs. \ref{Activated}(a,b,c,d) we plot the transport thermoelectric properties for different values of the ratio $V_{1}/V_{2}: a)1.2; \hspace{0.1cm} b)1.0; \hspace{0.1cm} c) 0.8; \hspace{0.1cm} d) 0.5$. Fig. \ref{Activated}(a) shows a high $ZT$ value, but the Power factor is very low, which limits the usefulness of this region. At the topological transition $V_{1}/V_{2}=1.0$ (Fig. \ref{Activated}(b)), $ZT$, $\xi$ and the $PF$ attain robust values. As $V_{1}/V_{2}$ decreases, the gap increases as indicated in Fig. \ref{Gap}; the $ZT$, $\xi$ and the $PF$ attain high values for $V_{1}/V_{2}=0.95$ (Fig. \ref{Activated}(c)). For $V_{1}/V_{2}=0.80$, the peak of $ZT$, $\xi$ and the $PF$ occur at around $T/V_{2}=0.1$ (Fig. \ref{Activated}(d)). However, the thermoelectric properties value tends to decrease for low values of $V_{1}/V_{2}$.

\section{Conclusions and Perspectives}
\label{sec8}

Topological insulating chains have many exciting properties.  These chains can be realized in materials with hybridized $sp$-states where the anti-symmetric nature of the hybridization between orbitals of different parities guarantees their topological properties.  We consider in this work monoatomic and diatomic $sp$-chains that map directly in the SSH and Rice-Mele problems, respectively. We obtain  the  density of states at the edge of a semi-infinite chain, which varies according to the topological phase of the chain.
We show that the weight of the zero energy modes in the non-trivial topological phase  vanishes continuously with the distance to the topological transition. 
In order to study the transport properties of the chains, we considered a simple device consisting of a quantum dot connected to two identical  semi-infinite $sp$  or Rice-Mele chains. Away from the topological transition and  at $T=0$, the current through the device vanishes since the chains  are insulators in their bulk, whether they are in the topologically trivial or non-trivial phases.  However, at finite temperatures there is activated transport that is different in the trivial and topological phases.

At the topological transition of the monoatomic, or SSH chains, and zero temperature, the conductance in the device has a finite  universal value $G/G_0 = 1$, independent of the parameters of the model like the coupling between the chains and the dot, as long as the energy of the dot $E_0=0$. Since, at the transition, the surface modes penetrate into the bulk,  the system carries current even at $T=0$. The normalized Wiedemann-Franz ratio turns out to be equal unity in terms of the Lorenz number. The thermal conductivity vanishes at $T=0$ even at the topological transition and  the thermopower of the monoatomic chains always vanishes due to particle-hole symmetry.

A different behavior arises when we consider diatomic $sp$-chains with different sub-lattices local energies. In this case the finite local energies break the chiral symmetry of the SSH Hamiltonian and the chain is now modeled by the Rice-Mele Hamiltonian. This system still presents non-trivial topological phases that are now characterized by Chern numbers. Interestingly,  the zero temperature dimensionless conductance  at the topological transition assumes  the value $G/G_0=1/2$, as would be expected for carriers with  a fractional charge and  is a consequence of the breakdown of chiral symmetry of the SSH model. The thermopower of the device in this case has an anomalous behavior and does not vanish at low temperatures.  It attains a universal value at $T=0$ consistent with  the result for the  conductance that implies fractional charges $q^*=1/2$ flowing in the system.  This is due to the double degeneracy of the system associated with the presence of zero energy modes. it is interesting to compare the physical properties of the fractional charge carriers in the topological insulators with those of Majorana modes in p-wave superconductors~\cite{Semenoff_2007}. 

Notice that the antisymmetric hybridization responsible for the non-trivial topological properties of chains does not mix the spins of the carriers. This is quite distinct from the case of spin-orbit interactions that mixes the spins. The consequence is that it is much easier to produce a singly polarized material  in the former case. 

The $sp$-chains, with edge modes in their topological phases are easier to realize in practice than $p$-wave superconductors. Carbyne, the one-dimensional allotropic form of carbon with  hybridized $sp$ orbitals provides a realization of these chains.  They are potentially useful systems exhibiting  properties that can be explored in a large temperature range. In particular, we show that varying $V_{1}/V_{2}$, the figure of merit and power factors can attain high values at high temperatures, making the system very attractive to be explored in technological applications.

\section{Appendix}
\label{appendix}

\subsection{SSH and sp-chain models}

The Hamiltonian describing the monoatomic sp-chain is given by~\cite{Foo1974, Continentino2014c},
\begin{eqnarray}
\label{sp}
\mathcal{H}_{sp}&=& \epsilon^0_s \sum_j c^{\dagger}_j c_j\! +\! \epsilon^0_p \sum_j p^{\dagger}_j p_j\! -\! \sum_j t_s (c^{\dagger}_j c_{j+1}  
\!+\! c^{\dagger}_{j+1} c_{j}) \nonumber  \\
&+& \! \sum_j t_p (p^{\dagger}_j p_{j+1} \!+\! p^{\dagger}_{j+1} p_{j}) 
+ \! V \sum_j (c^{\dagger}_j p_{j+1}\! -\!c^{\dagger}_{j+1} p_{j}) \nonumber \\
&-& \! V^{*} \sum_j (p^{\dagger}_j c_{j+1} - p^{\dagger}_{j+1} c_{j})  
\end{eqnarray}
where $\epsilon^0_{s,p}$ are the centers of the $s$ and $p$ bands, respectively. The $t_{s,p}$ represent the hopping of {\it spinless} electrons to neighboring sites in the same orbital and $V$ the antisymmetric hybridization between $s$ and $p$ states in neighboring sites.  Due to the different parities of the orbital states, this hybridization is odd-parity, such that,  in momentum space $V(-k)=-V(k)$. Then the mixing term breaks the parity symmetry of the system in spite that the chain is centro-symmetric.  Notice that the spin-orbit coupling (SOC) also breaks parity symmetry ~\cite{Caldas2013}, but differently from  hybridization that mixes quasi-particles with the same  spin,   SOC mixes quasi-particles with opposite spins~\cite{Caldas2013}. For simplicity we consider  here the case of spinless fermions and take the chemical potential $\mu=0$. We considerer symmetric bands, such that, $\epsilon^0_s=-\epsilon^0_p=\epsilon$ and assume $t_s=t_p=V=t$. 

The frequency dependent  Green's function $G_{00}$ at the edge of the semi-infinite chain can be obtained as ~\cite{Foo1976a}, 
\begin{equation}
\label{G00SP}
G_{00}(\omega)=\frac{1}{\omega}\left[\tilde{\omega}^2-\tilde{\epsilon}^2+1\pm \sqrt{(\tilde{\omega}^2-\tilde{\epsilon}^2+1)^2-4 \tilde{\omega}^2}\right]
\end{equation}
where $\tilde{\omega}=\omega/2t$ and $\tilde{\epsilon}=\epsilon/2t$. Comparing Eqs.~\ref{G00SSH} and~\ref{sp}, we obtain the formal relation between the SSH and $sp$-chain models.
Notice the factor $(1/2)$ difference that appears in Eq.~\ref{G00SSH} since for the $sp$-chain the full bands correspond to two electrons per site. 

\subsection{The $sp$-hybrid diatomic chain}

\begin{figure}[tbh]
\begin{center}
   \includegraphics[clip,width=0.47\textwidth,angle=0.0] {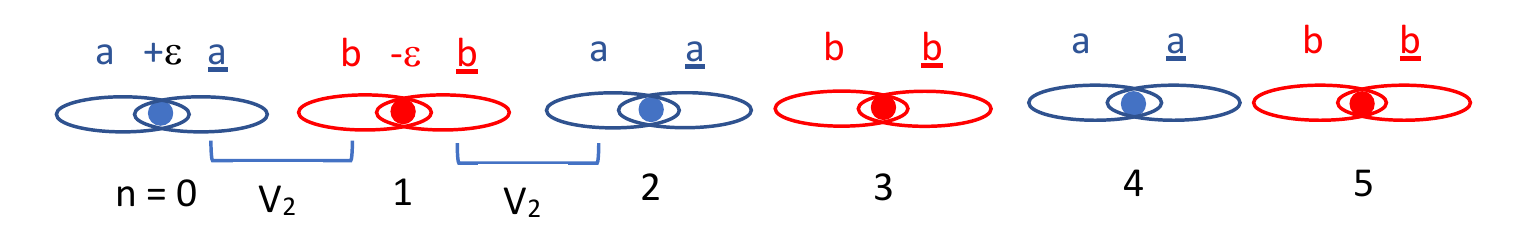}
\end{center}
    \caption{(Color online) Diatomic sp-chain with similar topological properties of the RM model,}
\label{diasp}
 \end{figure}

The Hamiltonian of the $sp$-hybrid, diatomic chain is given by~\cite{FooGian},
\begin{eqnarray}
H\!&=&\! \! \sum_n  \epsilon(n)(\alpha^{+}_n \alpha_n + \beta^+_n \beta_n)+ \! \sum_nV_1(n)(\alpha^+_n \beta_n + \beta^+_n \alpha_n) \nonumber \\
\!&+&\!\! V_2 \sum_n (\alpha^+_n \beta_{n+1}+ \beta^+_{n+1} \alpha_n)
\end{eqnarray}
where $\alpha^+_n$ creates an electron in orbital $a$ for $n$ even and in orbital $b$  for $n$ odd. The operator $\beta^+_n$ creates an electron in orbital $\underline{a}$ or $\underline{b}$ for $n$ even or odd respectively.  $\epsilon(n)$ takes values $+ \epsilon$ and $- \epsilon$ for even ($a$)  and odd ($b$) sites, respectively, as shown in Fig.~\ref{diasp}. Here we take $V_1(n)=V_1$ independent of the site to obtain a Hamiltonian similar to the Rice-Mele model.  
The constant terms $V_2$ connect orbitals $\underline{a}-b$ and $\underline{b}-a$ in different sites.

\section{Acknowledgments}

MAC acknowledges the Brazilian agencies CNPq, and Funda\c{c}\~ao de Amparo a Pesquisa do Estado do Rio de Janeiro FAPERJ for partial financial support. M.~S.~F. acknowledges financial support from the Brazilian National Council for Scientific and Technological Development (CNPq) Grant. Nr. 311980/2021-0 and to
Foundation for Support of Research in the State of Rio de Janeiro (FAPERJ) process Nr. 210 355/2018.

\bibliography{mucio}

\end{document}